\documentclass[aps,pre,twocolumn,superscriptaddress,showpacs,showkeys]{revtex4-1}
\usepackage{graphicx} 
\usepackage{bm}
\usepackage{amsmath}
\usepackage{amssymb}
\usepackage{times}
\renewcommand{\d}{\mathrm{d}}

\newcommand{\be}{\begin{equation}}
\newcommand{\ee}{\end{equation}}
\newcommand{\bea}{\begin{eqnarray}}
\newcommand{\eea}{\end{eqnarray}}
\newcommand{\bse}{\begin{subequations}}
\newcommand{\ese}{\end{subequations}}

\newcommand{\pf}{k_{\mathrm F}}
\newcommand{\kf}{k_{\mathrm F}}
\newcommand{\vf}{v_{\mathrm F}}
\newcommand{\ef}{\varepsilon_{\mathrm F}}

\newcommand{\q}{q}
\newcommand{\mathsym}[1]{{}}

\begin{document}

\title{Theory of superfluidity and drag force in the one-dimensional Bose gas}

\author{Alexander Yu.~Cherny}
\affiliation{Bogoliubov Laboratory of Theoretical Physics, Joint Institute for Nuclear
Research, 141980, Dubna, Moscow region, Russia}

\author{Jean-S\'ebastien Caux}
\affiliation{Institute for Theoretical Physics, Science Park 904, University of Amsterdam, 1090 GL
Amsterdam, The Netherlands}

\author{Joachim Brand}
\affiliation{Centre for Theoretical Chemistry and Physics and New Zealand Institute for Advanced Study, Massey University, Private Bag 102904 North Shore, Auckland 0745, New Zealand}

\date{\today}

\begin{abstract}
The one-dimensional Bose gas is an unusual superfluid. In contrast to higher spatial dimensions, the existence of non-classical rotational inertia is not directly linked to the dissipationless motion of infinitesimal impurities. Recently, experimental tests with ultracold atoms have begun and quantitative predictions for the drag force experienced by moving obstacles have become available. This topical review discusses the drag force obtained from linear response theory in relation to Landau's criterion of superfluidity. Based upon improved analytical and numerical understanding of the dynamical structure factor, results for different obstacle potentials are obtained, including single impurities, optical lattices and random potentials generated from speckle patterns. The dynamical breakdown  of superfluidity in random potentials is discussed in relation to Anderson localization and the predicted superfluid-insulator transition in these systems.
\end{abstract}


\maketitle

\tableofcontents

\section{Introduction}
\label{sec:intro}

Superfluidity in a neutral gas or liquid is not easily defined but rather understood as  a complex cluster of phenomena.
The associated properties may include
frictionless flow through thin capillaries, suppression of the classical inertial moment, metastable currents, quantized circulation (vortices), Josephson effect (coherent tunneling), and so on (see, e.g., Refs.~\cite{leggett99,leggett01:rev,sonin82,nozieres90:book,pitaevskii03:book,leggett06:book}). There is a close and deep analogy between superfluidity in a neutral system and superconductivity in a charged system \cite{nozieres90:book,leggett06:book}.

The 3D weakly-interacting Bose gas has all the superfluid properties mentioned above, which can be inferred from the existence of an order parameter represented by the wave function of the Bose-Einstein condensate (BEC). By contrast, there is no BEC in a repulsive 1D Bose gas even at zero temperature in the thermodynamic limit, provided interactions are independent of particle velocities \cite{bogoliubov61:book,hohenberg67}. This can be easily proved using the Bogoliubov ``$1/q^2$" theorem \cite{bogoliubov61:book}. This predicts a $1/q^2$ divergence at small momentum in the average occupation number $n_{q}$ for nonzero temperature and $1/q$ divergence for zero temperature. Nevertheless, the existence of BEC is neither a sufficient nor necessary condition for superfluidity \cite{leggett99,leggett06:book}, and a one-dimensional system of bosons may be superfluid under some conditions. However, whether a system is superfluid or not depends very much on how superfluidity is defined, because one-dimensional systems may exhibit only some but not all of the superfluid phenomena.

Here we study superfluidity in an atomic gas of repulsive spinless bosons in the 1D regime of very narrow ring confinement. The investigations focus mainly on the metastability of the circulating-current states in various regimes. However,  we also discuss another important aspect of superfluidity relevant to a 1D system, which is the non-classical moment of inertia or Hess-Fairbank effect \cite{1967_Hess_PRL_67} and the quantization of circulation. As we argue below, a perfect Hess-Fairbank effect and quantization of circulation occur for the homogeneous gas of repulsive spinless bosons in one dimension, while metastability of currents does not, in general. Note that the Hess-Fairbank effect is much easier to investigate than metastability of current because of its ``equilibrium" nature \cite{leggett73,leggett06:book}. Indeed, it can be explained with the properties of the low-lying energy excitation spectrum of the system due to the ability of the system to relax to the ground state in the reference frame where the walls (i.e.\ the trapping potentials) are at rest (see Sec.~\ref{sec:landau}). The metastability of currents is a much more complicated phenomenon to study, because at sufficiently large gas velocities, the system is obviously not in the ground state but in a metastable state. In order to study such an effect, one needs to understand transitions between states, which presents a more intricate problem.

Ideally, in order to study the decay of ring currents in a controlled manner, the gas should be kept in a ring or a torus-like geometry with defined defects. The defects may cause transitions to the states of lower energies, thus leading to energy dissipation, related to a friction force; this is called the drag force. The question of metastability then becomes equivalent to the drag force of a small and heavy impurity that is dragged through the resting gas.
In the scope of this topical review, we consider mainly infinitesimal impurities and calculate the lowest order terms in linear response of the interacting gas to perturbation by the impurity. The authors of \cite{sykes09} followed a different approach by calculating the effects of finite impurities on the flow of a weakly-interacting Bose-Einstein condensate.

In spite of rapid progress in experimental techniques along this line \cite{gupta05:ring_BEC,ryu07,olson07}, so far no conclusive experimental data on the drag force or metastability of ring currents in the 1D Bose gas is available, and thus, this is one of the outstanding fundamental questions remaining about the properties of ultra-cold Bose gases \cite{leggett01:rev}. Not long ago, an experiment along this line was carried out \cite{palzer09}, in which the propagation of spin impurity atoms through a strongly interacting one-dimensional Bose gas was observed in a cigar-shaped geometry. The motion of the center-of-mass position of the wave packet is described fairly well by the drag force, calculated with the dynamic structure factor of the Bose gas in the regime of infinite boson interactions. In the recent experiment of Ref.~\cite{catani11}, the dynamics of \emph{light} impurities in a bath of bosonic atoms was investigated and the decay of breathing mode oscillations was observed. In another line of experiments, atoms were subjected to a moving optical lattice potential and the momentum transfer was measured \cite{fallani04,fertig05,mun07}. This implies that one can experimentally obtain the drag force of a specific external potential acting on the gas. Experiments were also done with ultra-cold atoms in random and pseudo-random potentials. The direct observation of Anderson localization was reported  in Refs.~\cite{billy08,roati08,deissler10}. In particular, spreading of a 1D Bose gas in artificially created random potentials was experimentally investigated \cite{billy08}. Below we show that the superfluid-insulator phase diagram of such a system can be obtained by calculating the drag force.

The notion of drag force turns out to be theoretically fundamental, because it generalizes Landau's famous criterion of superfluidity. According to Landau, an obstacle in a gas, moving with velocity $v$,  may cause transitions from the ground state of the gas to excited states lying on the line $\varepsilon=p v$ in the energy-momentum space. If all the spectrum is above this line, the motion cannot excite the system, and it is thus superfluid. However, it is also possible that even when the line intersects the spectrum, the \textit{transition probabilities} to these states are strongly suppressed due to boson interactions or to the specific kind of external perturbing potential. In this case, the drag force gives us a \emph{quantitative} measure of superfluidity.

This paper is organized as follows. The basic model of the 1D Bose gas considered in this paper is introduced in Sec.~\ref{sec:model}. In the subsequent section, we study the Hess-Fairbank effect and its relation to the Landau criterion of superfluidity. In Sec.~\ref{sec:general}, we derive an expression for the drag force through the dynamic structure factor within linear response theory and show that the notion of the drag force generalizes the Landau criterion of superfluidity. The values of the drag force in various limiting regimes are calculated in Sec.~\ref{sec:diffregimes}. In the subsequent section, the Luttinger liquid theory is exploited to describe the drag force at small impurity velocities. An exact method for obtaining the drag force for a finite number of bosons, exploiting the algebraic Bethe ansatz, is briefly discussed in Sec.~\ref{sec:ABA}. In Sec.~\ref{sec:effapprox} we consider a simple interpolation formula for the DSF, which works for arbitrary strength of the interparticle interactions. In the subsequent section, we show that another approach to the one-dimensional Bose gas, the theory of phase slip transitions \cite{kashurnikov96}, reproduces the same power-law behaviour for the drag force as the Luttinger liquid theory does. In Sec.~\ref{sec:beyond_linear}, the drag force in the limit of infinite interactions is obtained analytically beyond linear response theory. The decay of ring currents is examined in Sec.~\ref{sec:velocity_damping}. In Sec.~\ref{sec:shallow_lattice}, the zero-temperature phase diagram is obtained for the superfluid-insulator transition of the one-dimensional Bose gas in moving shallow lattices for arbitrary values of velocity, filling factor, and strength of boson interactions. The motion of the 1D Bose gas in random potentials and the related phase diagram at zero temperature are considered in Sec.~\ref{sec:random}. Finally, in the conclusion the results and prospects are briefly discussed.

\section{Landau criterion of superfluidity and Hess-Fairbank effect}
\label{sec:landau}\label{sec:model}

{\em Model} -- We model cold bosonic atoms in a waveguide-like micro trap by a simple 1D gas of $N$ bosons with point interactions of strength $g_\mathrm{B}>0 $ ({\it i.e.} the Lieb-Liniger (LL) model \cite{lieb63:1})
\begin{equation}
H =  \sum_{i=1}^N -\frac{\hbar^2}{2 m}\frac{\partial^2}{\partial x_i^2}
+ g_{\text{B}} \sum_{1\leqslant i<j\leqslant N} \delta(x_i - x_j)
\label{LLham}
\end{equation}
and impose periodic boundary conditions on the wave functions. The strength of interactions can be measured in terms of the dimensionless parameter $\gamma\equiv m g_{\text{B}}/(\hbar^2 n)$, where $n$ is the linear density and $m$ is the mass. In the limit of large $\gamma$, the model is known as the Tonks-Girardeau (TG) gas \cite{1936_Tonks_PR_50,1960_Girardeau_JMP_1}. In this limit, it can be mapped onto an ideal \emph{Fermi} gas since infinite contact repulsions emulate the Pauli principle. In the opposite limit of small $\gamma$, we recover the Bogoliubov model of weakly interacting bosons. For an overview of theoretical approaches to the one-dimensional Bose gas see the recent review \cite{cazalilla11}.

\begin{figure}[tbh]
\includegraphics[width=.9\columnwidth]{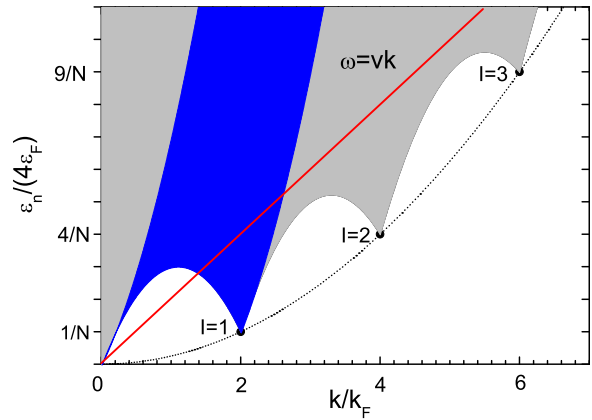}
\caption{\label{fig:schem} (Color online) Schematic of the excitation spectrum of the 1D Bose gas in a perfectly isotropic ring. The supercurrent states $I$ lie on the parabola $\hbar^2k^2/(2 N m)$ (dotted line). Excitations occur in the shaded area; the discrete structure of the spectrum is not shown for simplicity. The blue (dark) area represents single particle-hole excitations \cite{lieb63:2}. Motion of the impurity with respect to the gas causes transitions from the ground state to the states lying on the straight (red) line. Reproduced from Ref.~\cite{cherny09a}, Copyright \copyright 2009 The American Physical Society}
\end{figure}

{\em Landau criterion} --
In the LL model the total momentum is a good quantum number \footnote{In this paper we use the linear momentum and coordinates and velocities. The angular momentum and angle and angular velocity can easily be written as $L_z= p L/2\pi$, $\varphi=2\pi x/L$, $\omega_z=2\pi v/L$, respectively.},
and periodic boundary conditions quantize it in units of $2\pi\hbar/L$, where $L$ is the ring circumference. Classification of all the excitations can be done \cite{lieb63:2} in the same manner as for the Tonks-Girardeau gas of non-interacting fermions ($\gamma\to+\infty$). Thus, in order to create an elementary particle-like excitation, one needs to add a quasimomentum beyond the occupied Fermi segment. By contrast, for a hole-like excitation, one needs to remove a quasimomentum lying inside the Fermi segment. All the excitations can be constructed from the above elementary excitations. Due to the conservation of total number of particles, the number of particle-like excitations coincides with that of hole-like excitations. The low-lying spectrum of $N=nL$ bosons as shown in Fig.~\ref{fig:schem} has local minima \cite{haldane81} at the supercurrent states $I$ ($I=0,1,2,\ldots$) with momenta $p_I = 2 \pi n \hbar I$ and excitation energies
\begin{equation}
\varepsilon_I = p_I^2/(2 N m).
\label{EI}
\end{equation}
These correspond to Galilean transformations of the ground state with velocities $v_I =
p_I/(N m)$. The minima do not depend on interactions and tend to zero in the limit of
large system size at constant density. The first supercurrent state is also called the
umklapp excitation \cite{lieb63:1} by analogy with periodic lattices
because it can be reached from the ground state by imparting the
momentum $\hbar K_\mathrm{r}$ to each particle,
where  $K_\mathrm{r}=2\pi/L$ is the reciprocal wave vector in the ring geometry. As explained above, this changes the total momentum while preserving the internal state of the system.

Suppose that the gas is put into rotation with linear velocity $v_I$, and after that, is braked with an artificial macroscopic ``obstacle,'' {\it e.g.}, created by a laser beam \cite{raman99:critical_velocity}. In the reference frame where the gas is at rest, the obstacle moves with velocity $v_I$. In a superfluid we expect to see no energy dissipation, and the drag force is zero (the current is persistent). Otherwise one can observe decay of the current. It follows from energy conservation that the transitions from the ground state caused by the moving obstacle with velocity $v$, lie on the line $\omega= v k$ in the energy-momentum plane.  According to Landau, if the excitation spectrum lies above this line, the motion cannot excite the system, which is then superfluid. As is seen from Fig.~\ref{fig:schem}, the Landau critical velocity (when the line touches the spectrum) equals $v_\mathrm{c}=v_1/2=\hbar\pi/(m L)$. This implies that any supercurrent state with $I\geqslant1$ is unstable since $v_I>v_\mathrm{c}$. However, in 3D we also have similar supercurrent states, which apparently leads to the paradoxical absence of current metastability. The solution to this is that we need to consider not only the spectrum but also \emph{probabilities} of excitations. Below we argue that in the 3D case, the probability to excite supercurrents is vanishingly small, while in the 1D case it depends on the strength of bosonic interactions.

\begin{figure}[tbhp]
\centerline{\includegraphics[width=.9\columnwidth]{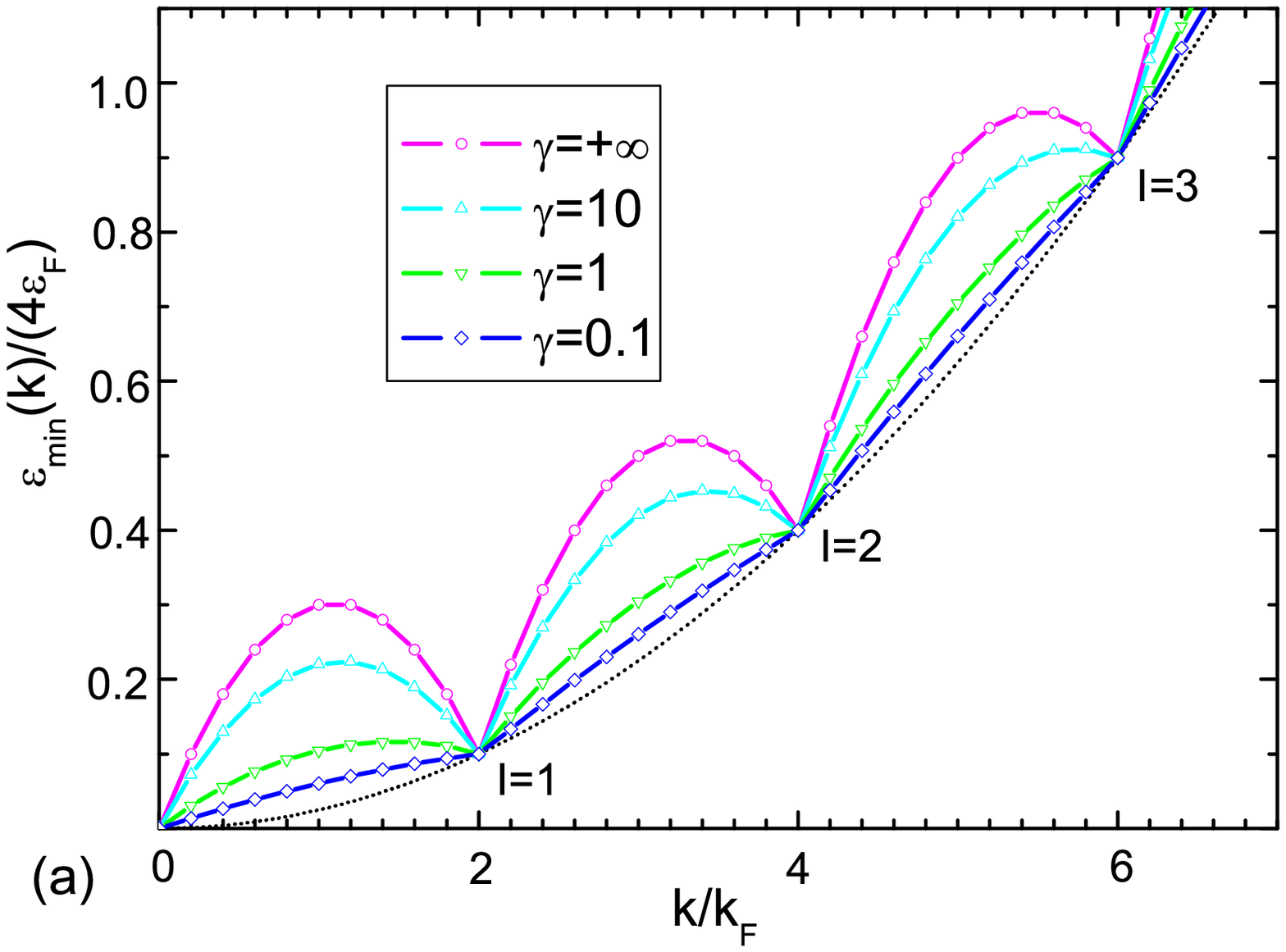}}
\centerline{\includegraphics[width=.9\columnwidth]{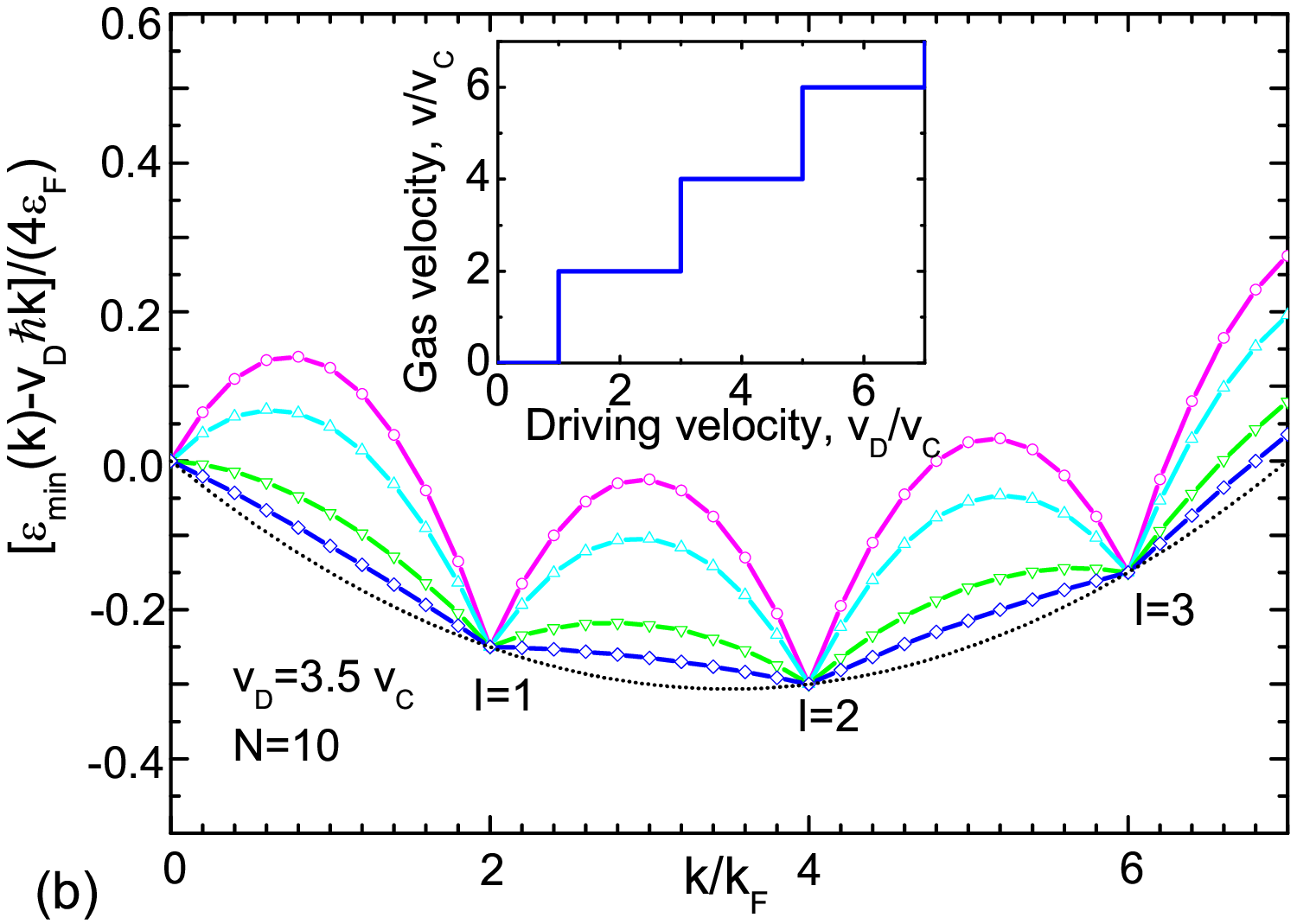}}
\caption{\label{fig:spectrum}(Color online) (a) Low-lying excitation spectrum $\varepsilon_\mathrm{min }(k)$ for $N=10$ particles for 1D repulsive bosons. At $\gamma\ll 1$, the low-lying spectrum verges towards that of the ideal Bose gas, which lies on the straight segments between points $I=1,2,3,\ldots$. (b) Quantization of current velocity for 1D repulsive bosons under influence of a moving trap. Shown are the low-energy excitations of the 1D Bose gas in the moving frame $\varepsilon_\mathrm{min}(k) - v_\mathrm{D} \hbar k$, calculated from the Bethe-ansatz equations \cite{lieb63:1} for different values of the coupling strength (compare with $\varepsilon_\mathrm{min }(k)$ of Fig.~\ref{fig:schem}). Inset: The velocity of the gas at equilibrium changes abruptly at values of driving velocity $v_\mathrm{D}/v_\mathrm{c}=1,3,5,\ldots$, since the gas occupies the state with lowest energy. In particular, the system is at rest when the driving velocity is less than $v_\mathrm{c}=\vf/N$ (perfect Hess-Fairbank effect).  Here, $\kf\equiv\pi n$, $\vf=\hbar\kf/m$, and $\vf\equiv\hbar^2 \kf^2 /(2 m)$. Reproduced from Ref.~\cite{cherny10a}, Copyright \copyright\ Siberian Federal University.}
\end{figure}

{\em Hess-Fairbank effect} --
When the walls of a toroidal container are set into rotation adiabatically with a small tangential velocity $v_\mathrm{D}$, a superfluid stays at rest while a normal fluid follows the container. This effect leads to a nonclassical rotational inertia of superfluid systems, which can be used to determine the superfluid fraction \cite{leggett73,leggett06:book}. For the 1D Bose gas, rotation of the annular trap amounts to shifting the excitation spectrum to $\varepsilon - v_\mathrm{D} p$ as shown in Fig.~\ref{fig:spectrum}. It is assumed that an unspecified relaxation mechanism allows the system to relax to the ground state in the frame where {\it the trap is at rest}. The low-lying LL excitation spectrum is a convex function of momentum for $0\leqslant p\leqslant p_1$ \cite{lieb63:2}, and, hence, the momentum zero state remains the ground state for $|v_\mathrm{D}| <v_c$. This leads to the Hess-Fairbank effect for the 1D Bose gas for arbitrary repulsive interactions $\gamma >0$ \cite{ueda99,cazalilla04}. According to this {\it equilibrium} property which is completely determined by the low-lying energy spectrum \cite{leggett99}, the 1D Bose gas has a 100\% superfluid fraction and zero rotational inertia at zero temperature. The same results were obtained by using instanton techniques \cite{buchler03} and within Luttinger liquid theory for a finite number of bosons \cite{cazalilla04,citro09}.

It is the convexity of the low-lying excitation spectrum between the supercurrent states (see Fig.~\ref{fig:spectrum}) that allows us to obtain this result without numerical calculations. The minima of energy can be reached only in the supercurrent states (see Fig.~\ref{fig:spectrum}b), whose energy are known analytically with Eq.~(\ref{EI}). In the usual way (see, e.g., Ref.~\cite{cazalilla04}), the superfluid component can be defined and obtained numerically through the second derivative of the phase $\varphi$ in  the \emph{twisted} boundary conditions $\psi(x_1, \ldots,x_j+L, \ldots,x_N) = exp(i\varphi) \psi(x_1, \ldots,x_j,\ldots,x_N)$, $j=1,\ldots,N$, imposed on the wave functions. We emphasize that this procedure is equivalent to the method used in this paper, because the Galilean transformation implies a phase gradient of the wave functions observed in the frame of the moving walls, and the gradient leads to twisted boundary conditions. It is also consistent with the definition of the superfluid component in three dimensions through the Fourier-transformed autocorrelation function of the transverse current \cite{leggett73,leggett06:book}.

Note that the Bose-Fermi mapping for the TG gas implies that periodic boundary conditions for the Bose gas only translate into periodic boundary conditions for the Fermi gas for odd $N$ but into antiperiodic ones for even $N$ \cite{cheon99}, in contrast to the true Fermi gas with periodic conditions. One can see that for even $N$, there is no Hess-Fairbank effect in the true Fermi gas (although the TG gas always shows the Hess-Fairbank effect). This is due to instability of fermions at the Fermi point, which results from the degeneracy of the ground state for an even number of fermions.

\section{Drag force as a generalization of Landau's criterion of superfluidity}
\label{sec:general}

In order to study frictionless motion, let us consider an impurity of mass $m_\mathrm{i}$, moving with velocity $\bm{v}$ in a medium of particles.
For the major part of this topical review, we will treat the interaction between the impurity and Bose gas perturbatively in linear response theory and thus reduce the problem to calculating properties of the integrable 1D Bose gas, although in  general the system will lose integrability when adding an impurity.
Using Fermi's golden rule, one can easily show (see App.~\ref{sec:dffgr}) that the resulting friction leads to an energy loss per unit time given by
\begin{equation}
\dot{E}=-\!\int\!\frac{\d^D q}{(2\pi)^{D-1}}|\widetilde{V}_\mathrm{i}(q)|^2 n
\frac{(\bm{q}\cdot\bm{v}\!-\!\frac{\hbar q^2}{2 m_\mathrm{i}})
S(q,\bm{q}\cdot\bm{v}\!-\!\frac{\hbar q^2}{2 m_\mathrm{i}})}{N}.
\label{enloss}
\end{equation}
Here $\widetilde{V}_\mathrm{i}(q)$ is the Fourier transform of the interaction potential $V_\mathrm{i}(r)$ between the impurity and the particles, $D$ is the spatial dimension, $n$ is the density of particles (number of particles per $D$-dimensional volume), and $S(q,\omega)$ is the dynamical structure factor (DSF) of the medium. It is given by the definition \cite{pitaevskii03:book}
\begin{equation}
S(q,\omega)={\cal Z}^{-1}\sum_{n,m}e^{-\beta E_{m}}|\langle m|\delta\hat{\rho}_{\bm{q}}|n
\rangle|^{2} \delta(\hbar \omega-E_{n}+E_{m}),
\label{sqomega}
\end{equation}
with ${\cal Z}=\sum_{m}\exp(-\beta E_{m})$ being the partition function and $\beta$ being the inverse temperature. Here $\delta\hat{\rho}_{\bm{q}}=\sum_{j}e^{-i \bm{q}\cdot\bm{r}_j}-N\Delta(\bm{q})$ is the Fourier component of the operator of the density fluctuations, $\Delta(\bm{q})=1$ at $\bm{q}=0$ and $\Delta(\bm{q})=0$ otherwise. At zero temperature, the structure factor takes a simpler form
\begin{equation}
S(q,\omega)=\sum_{n}|\langle 0|\delta\hat{\rho}_{\bm{q}}|n \rangle|^{2}
\delta(\hbar \omega-E_{n}+E_{0}).
\label{sqomega0}
\end{equation}
The DSF relates to the time-dependent density correlator through the Fourier transformation
\begin{equation}
S(q,\omega)=N\int \frac{\d t\d^{D}r}{(2\pi)^D\hbar}\,e^{i(\omega t-\bm{q}\cdot\bm{r})}
\langle\delta\hat{\rho}(\bm{r},t)\delta\hat{\rho}(0,0)\rangle/n,
\label{densdens}
\end{equation}
where $\delta\hat{\rho}(\bm{r},t)\equiv\sum_{j}\delta(\bm{r}-\bm{r}_j(t))-n$ is the operator of the density
fluctuations. The DSF obeys the $f$-sum rule \cite{pitaevskii03:book}
\begin{equation}\label{fsumrule}
\int_{-\infty}^{+\infty}\d\omega\, \omega S(q,\omega)= N{q^{2}}/{(2m)}.
\end{equation}

The drag force is defined by the formula $\dot{E}=-\bm{F}_\mathrm{v}\cdot\bm{v}$. In this paper, we will use the expression (\ref{enloss}) for a heavy  impurity  $v\gg \hbar q/m_\mathrm{i}$ in one dimension. It yields for the drag force $F_{\mathrm{v}} = \int_{-\infty}^{+\infty}\d q\,q\,|\widetilde{V}_\mathrm{i}(q)|^2 S(q,q v)/L$. Using the properties of the DSF $S(q,\omega)= S(-q,\omega)= e^{\beta\hbar\omega} S(q,-\omega)$, which follow  from its definition (\ref{sqomega}), we obtain
\begin{equation}
F_{\mathrm{v}}=\int_{0}^{+\infty}\d q\,q\,\,|\widetilde{V}_\mathrm{i}(q)|^2
S(q,q v)[1-\exp(-\beta \hbar q v)]/L.
\label{dragf1D}
\end{equation}
This is the most general form of the drag force within linear response theory. The form of the impurity interaction potential can be of importance as, e.g., in the cases of shallow lattices and random potentials discussed in Secs.~\ref{sec:shallow_lattice} and \ref{sec:random}, respectively. When the impurity interaction is of short-range type, we can replace it with good accuracy by a contact interaction $V_{\mathrm{i}}(r) =g_{\mathrm{i}}\delta(x)$, which leads to $\widetilde{V}_\mathrm{i}(q)=g_{\mathrm{i}}$ and yields at zero temperature \cite{astrakharchik04}
\begin{equation}
F_{\mathrm{v}} \equiv\frac{2g_{\mathrm{i}}^{2}nm}{\hbar^2}f_\mathrm{v}
=g_{\mathrm{i}}^{2}\int_{0}^{+\infty}\d q\,q S(q,q v)/L. \label{dragf1Dpoint}
\end{equation}
In the first equality, we introduce the dimensionless drag force $f_\mathrm{v}$ with the help of the ``natural" unit ${2g_{\mathrm{i}}^{2}nm}/{\hbar^2}$. Its physical nature will be discussed below in Secs.~\ref{sec:largev} and \ref{sec:beyond_linear}.

The notion of drag force generalizes the Landau criterion of superfluidity. Indeed, the integral in Eq.~(\ref{dragf1Dpoint}) is taken along the line $\omega=q v$ in the $\omega$-$q$ plane. If the excitation spectrum lies completely above the line then the integral vanishes, as one can see from the DSF definition (\ref{sqomega0}). On the other hand, the integral can be infinitesimally small or vanish even if the spectrum lies below the line but the excitation probabilities, given by the corresponding matrix elements $\widetilde{V}_\mathrm{i}(q) \langle 0|\delta \hat{\rho}_{\bm{q}}|n \rangle$, are suppressed. The drag force can vanish even if the system is not superfluid in principle, but excitations are not accessed with a given potential.
This happens when the matrix elements $\langle 0|\delta\hat{\rho}_{\bm{q}}|n \rangle$ are not small but the Fourier transform of the impurity potential $\widetilde{V}_\mathrm{i}(q)$ takes non-zero values only in a finite region of $q$-space. We consider such an interesting case in Secs.~\ref{sec:shallow_lattice} and \ref{sec:random} below.

Thus, determining the drag force within linear response theory is reduced to the problem of calculating the dynamic structure factor. Below we consider various approximations for the dynamic structure factor and the associated drag force.

\section{The drag force in different regimes}
\label{sec:diffregimes}

\subsection{Large impurity velocities}
\label{sec:largev}

Let us consider the case of contact interactions with the impurity, $V_{\mathrm{i}}(r) =g_{\mathrm{i}}\delta(x)$. At large impurity velocity $v \gg \hbar\pi n/m$, the main contribution to the integral in Eq.~(\ref{dragf1Dpoint}) comes from the high momentum region of the DSF, which can be calculated analytically \cite{hohenberg66}. Indeed, at large velocities, the momentum transfer from the impurity to the particles is big enough to neglect the interparticle interactions. Then one can use the DSF values of the ideal gas \cite{hohenberg66,pitaevskii03:book}
\begin{align}
S(q,\omega)&=\sum_{p}n_p(1\pm
n_{p+q})\delta\bigg(\hbar\omega-\frac{\hbar^2\q^2}{2m}-\frac{\hbar^2pq}{m}\bigg)\nonumber\\
&\simeq
\sum_{p}n_p\delta\bigg(\hbar\omega-\frac{\hbar^2\q^2}{2m}-\frac{\hbar^2pq}{m}\bigg).
\label{dsfid}
\end{align}
Here $n_p\equiv\langle a^{\dag}_p a_p\rangle$ is the average occupation numbers of particles, and the plus is taken for bosons and the minus for fermions. The second equality in Eq.~(\ref{dsfid}) is due to the large momentum $\hbar q\simeq mv \gg \hbar\pi n$, which leads to $n_p n_{p+q}\simeq 0$ for arbitrary values of $p$. Substituting Eq.~(\ref{dsfid}) into Eq.~(\ref{dragf1Dpoint}) yields the value of the drag force ${2g_\mathrm{i}^2 m n}/{\hbar^2}$, which can be used as the force unit.

It is not difficult to see that this result is valid \emph{beyond} linear response theory, given by Eq.~(\ref{dragf1Dpoint}). Indeed, at sufficiently large particle velocity, the initial particle momentum can be neglected during the scattering. Then in the reference frame where the impurity remains at rest, the relative particle momentum is $\hbar q\simeq mv$. Therefore, the reflection coefficient is determined by the squared absolute value of the scattering amplitude in the Born approximation $m^2g_\mathrm{i}^2/\hbar^4q^2$. Each particle, being reflected, transfers momentum $2\hbar q$. The total number of scattered particles per unit time is  $nv=n\hbar q/m$. The product of the last three quantities yields the value of momentum transfer per unit time, that is, the drag force. This quantity, independent of the wave vector, is nothing else but the limiting value of the drag force, obtained above.

\subsection{The Tonks-Girardeau regime}
\label{sec:TG}

\begin{figure}[tbhp]
\includegraphics[width=.8\columnwidth]{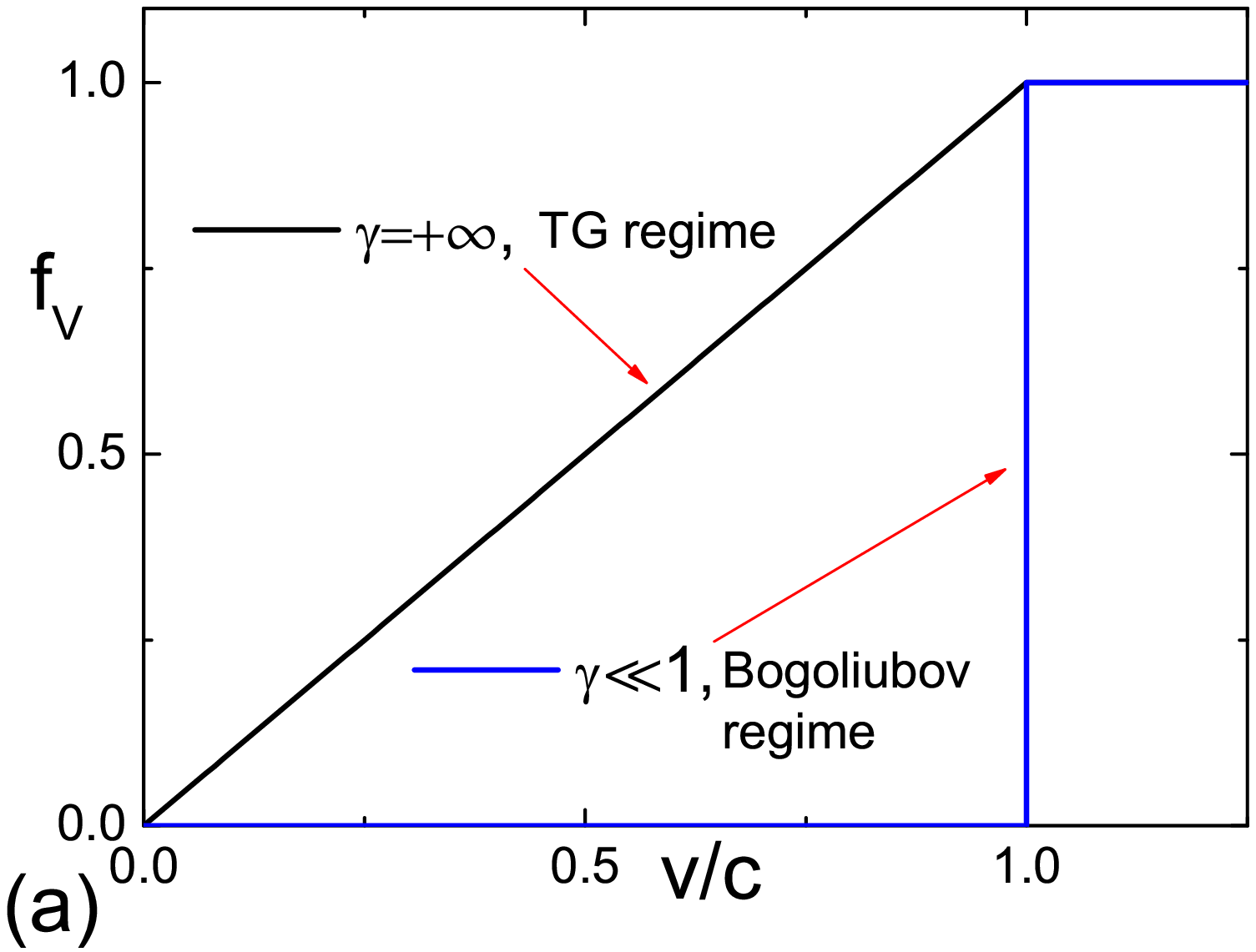}\\
\includegraphics[width=.8\columnwidth]{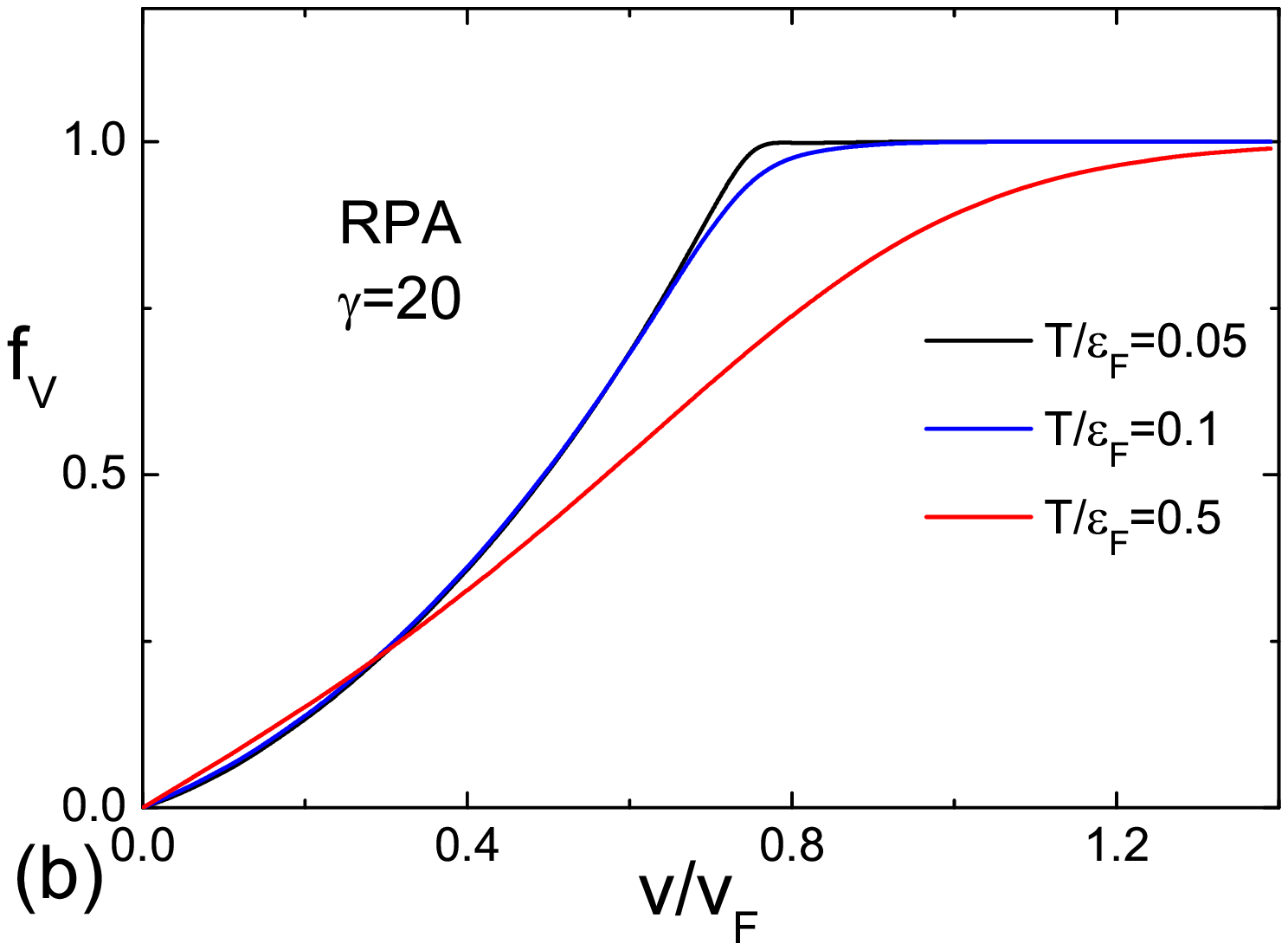}
\caption{\label{fig:DF_var_approx} (Color online) The dimensionless drag force (\ref{dragf1Dpoint}) versus impurity velocity in various approximations, $c$ is the speed of sound. (a) Tonks-Girardeau and Bogoliubov regimes. (b)  The random phase approximations (RPA) at non-zero temperatures. The curve becomes smoother when the temperature grows.}
\end{figure}

Let us first investigate the drag force for the TG gas ($\gamma\to+\infty$), having the same structure factor as the ideal Fermi gas. The DSF at zero temperature is well-known in the thermodynamic limit \cite{brand05,cherny06}
\begin{equation}
S(k,\omega)\frac{\ef}{N}= \frac{\kf}{4 k}
\label{DSFTG}
\end{equation}
for $\omega_{-}(k)\leqslant\omega\leqslant\omega_{+}(k)$, and zero otherwise. Here $\omega_\pm(k)={\hbar |2 \pf k \pm k^2|}/{(2 m)}$ are  the limiting dispersions that bound quasiparticle-quasihole excitations (see Fig.~\ref{fig:omplmi}). By definition, $\kf\equiv\pi n$ and $\ef\equiv\hbar^{2}\kf^{2}/(2m)$ are the Fermi wave vector and energy of the TG gas, respectively. As follows from Eq.~(\ref{DSFTG}) for the DSF, the transition probability from the ground state is inversely proportional to the momentum transfer but does not depend on the excitation energies within the borders $\omega_\pm(k)$. Simple integration in Eq.~(\ref{dragf1Dpoint}) then yields
\begin{equation}
f_\mathrm{v}=\left\{ \begin{array}{ll}
v/\vf,& 0\leqslant v\leqslant \vf,\\[1ex]
1,     & v\geqslant \vf,
\end{array}\right.
\label{dragfTG}
\end{equation}
where $\vf\equiv\hbar\pi n/m$ is the sound velocity in the TG regime. The result is represented in Fig.~\ref{fig:DF_var_approx}a. The TG gas is obviously not superfluid.

\subsection{The drag force in the Bogoliubov regime}
\label{sec:bogregime}

The opposite regime to the TG gas is the limit of weak interactions $\gamma\ll 1$ (the Bogoliubov regime). The crucial point in Bogoliubov theory \cite{bogoliubov47} is the developed Bose-Einstein condensate. In spite of the absence of Bose-Einstein condensation in one dimension \cite{book:bogoliubov70,hohenberg67}, the upper dispersion curve $\omega_{+}(k)$ is well described at small $\gamma$ \cite{lieb63:2} by the Bogoliubov relation
\begin{equation}
\hbar\omega_k=\sqrt{T_k^{2}+4 T_k \ef \gamma/\pi^{2}},
\label{bogdisp}
\end{equation}
where $T_{k}=\hbar^{2}k^{2}/(2m)$ is the free-particle energy spectrum. The Bogoliubov theory also yields the correct values of the ground state energy and chemical potential. This paradox can be explained \cite{cherny09} by a strong singularity of the DSF near $\omega_{+}$ in the Bogoliubov regime. As a result, it is localized almost completely within a small vicinity of the upper branch. This tendency can be seen even at $\gamma=10$ (Fig.~\ref{fig:omplmi}). Thus, the behavior of the DSF simulates the $\delta$-function spike, which appears due to the Bose-Einstein condensate. One can simply put $S_{\mathrm{Bog}}(k,\omega)=C \delta(\omega-\omega_{k})$ and determine the constant $C$ from the $f$-sum rule (\ref{fsumrule})
\begin{equation}
S_{\mathrm{Bog}}(k,\omega)=N\frac{T_{k}}{\hbar\omega_k} \delta(\omega-\omega_{k}).
\label{dsfbog}
\end{equation}

The drag force is then obtained analytically: it is a step function
\begin{equation}
f_\mathrm{v}=\left\{ \begin{array}{ll}
0,& 0\leqslant v\leqslant c,\\[1ex]
1,     & v\geqslant c,
\end{array}\right.
\label{gragfBog}
\end{equation}
shown in Fig.~\ref{fig:DF_var_approx}a. The sound velocity in the Bogoliubov regime is given by $c=\vf\sqrt{\gamma}/\pi$. At non-zero temperatures the result does not change. This result was first obtained in Ref.~\cite{astrakharchik04} by means of the GP equation. Note that the supersonic impurity motion in three dimensions was earlier studied by the same method in Ref.~\cite{kovrizhin01}.

\subsection{The linear approximation near the Tonks-Girardeau regime}
\label{sec:linear}

For finite $\gamma$, the model can also be mapped onto a Fermi gas \cite{cheon99} with local interactions, inversely proportional to $g_{\text{B}}$ \cite{girardeau04,granger04,brand05,cherny06}. Using the explicit form of the interactions, one can develop the time-dependent Hartree-Fock scheme \cite{brand05,cherny06} \emph{in the strong-coupling regime} with small parameter $1/\gamma$. Approximations of the linear response functions on this level are known as Random Phase approximation (RPA) with exchange or generalized RPA \cite{nozieres90:book,pines61:book}. The scheme yields the correct expansion of the DSF up to the first order \cite{brand05,cherny06}
\begin{equation}
S(k,\omega)\frac{\ef}{N}= \frac{\kf}{4 k}\left(1+\frac{8}{\gamma}\right)
+\frac{1}{2\gamma}\ln \frac{\omega^{2}-\omega_{-}^{2}} {\omega_{+}^{2}-\omega^{2}}+
O\left(\frac{1}{\gamma^2}\right),
\label{DSFlinear}
\end{equation}
for $\omega_{-}(k)\leqslant\omega\leqslant\omega_{+}(k)$, and zero elsewhere. The symbol $O(x)$ denotes terms of order $x$ or smaller. The limiting dispersions in the strong-coupling regime take the form
\begin{equation}
\omega_\pm(k)=\frac{\hbar |2 \pf k \pm k^2|}{2 m}\left(1-\frac{4}{\gamma}\right)
+O\left(\frac{1}{\gamma^2}\right).
\label{ompmstrong}
\end{equation}
From the linear part of the dispersion (\ref{ompmstrong}) at small momentum $\omega_\pm(k)\simeq ck$, we obtain the well-known result \cite{lieb63:2,cazalilla04} for the sound velocity at zero temperature $c=\vf(1-4/\gamma)+O(1/\gamma^2)$.

In the same manner as in Sec.~\ref{sec:TG}, we derive from Eqs.~(\ref{dragf1Dpoint}) and
(\ref{DSFlinear})
\begin{equation}
f_\mathrm{v}=\left\{ \begin{array}{ll}
\bigg[1+8\dfrac{\ln(v/c)}{\gamma}\bigg] \dfrac{v}{c},& 0\leqslant v\leqslant c ,\\[3ex]
1,     & v\geqslant c.
\end{array}\right.
\label{dragflinear}
\end{equation}

As expected, the linear approximation (\ref{dragflinear}) works badly for small values of impurity velocity. This is due to the anomalous  behaviour of the DSF within the linear approximation in vicinity of the umklapp excitation point at $\omega=0$, $k=2\kf$. Indeed, Eq.~(\ref{dragflinear}) leads to unphysical negative values of the drag force at sufficiently small impurity velocities. For this reason, we need a more careful examination of the drag force in this regime.

\section{Theoretical approaches}
\subsection{Random phase approximation near the Tonks-Girardeau regime}
\label{sec:RPA}

The DSF in the RPA  was calculated and described in details in Sec. IVB of Ref.~\cite{cherny06}.  The RPA is based on the Hartree-Fock scheme, which is appropriate at sufficiently large interactions between bosons $\gamma>8$. The equation for the drag force within the RPA is derived in App.~\ref{app:DF_RPA}.

The RPA approximation has some advantages. It always gives positive values of the drag force and is applicable at non-zero temperatures. The results are shown in Fig.~\ref{fig:DF_var_approx}b. The RPA scheme can be extended to non-homogeneous systems \emph{beyond the local density approximation} \cite{gattobigio06}. However, the RPA scheme also works badly in the vicinity of the umklapp excitation. As a consequence, the formula for the drag force (\ref{dragfsmallv1}), obtained within the RPA, does not reproduce the correct power-law behaviour at small velocities, see discussions in Secs.~\ref{sec:link_Luttinger} and \ref{sec:effapprox} below. Another disadvantage of the scheme is that the drag force as a function of velocity has an unphysical peak near $v=c$ at zero temperatures, which becomes quite pronounced for $\gamma\lesssim 15$. This is an apparent artifact of RPA. In the next sections we consider much better approximations that work for all values of the interaction strength and impurity velocities.

\subsection{Luttinger liquid theory}
\label{sec:link_Luttinger}

Luttinger liquid theory allows us to correctly describe the low-energy excitations of a 1D system of particles whose interactions are independent of the velocities. It faithfully handles the nonlinear effects of these interactions and yields values of the DSF \emph{at low energies} for arbitrary coupling strength \cite{haldane81,astrakharchik04} and thus can be used to calculate the drag force at small values of the impurity velocity.

The behaviour of the DSF in the vicinity of the umklapp excitation is related to the long-range asymptotics of the time-dependent density-density correlator. The asymptotics for $\kf x\gg 1$ and $t\leqslant x/c$
\begin{align}
\frac{\langle\delta\hat{\rho}(x,t)\delta\hat{\rho}(0,0)\rangle}{n^2}
\simeq &-\frac{K}{4\pi^2 n^2}\left(\frac{1}{(x-ct)^2}+ \frac{1}{(x+ct)^2}\right) \nonumber\\
&+ \widetilde{A}_1(K) \frac{\cos(2\kf x)}{n^{2K}(x^2-c^2 t^2)^{K}} \label{den_den_t}
\end{align}
can be justified by generalizing Haldane's method \cite{haldane81} or using perturbation theory \cite{popov72,popov83:book} or quantum inverse scattering method \cite{korepin93:book} or conformal field theory \cite{mironov91,gogolin98:book}. The first two terms are related to the behaviour of the DSF in vicinity of $\omega=0$, $q=0$, while the third term is related to the umklapp excitations. Substituting the third term of Eq.~(\ref{den_den_t}) into Eq.~(\ref{densdens}) yields the DSF in the vicinity of the ``umklapp" point ($k=2\kf=2\pi n$, $\omega =0$) \cite{astrakharchik04}
\begin{equation}
\frac{S(k,\omega)}{N}=\frac{n c}{\hbar\omega^{2}}
\left(\frac{\hbar\omega}{m c^{2}}\right)^{2K}
A_1(K)\left(1-\frac{\omega^{2}_{-}(k)}{\omega^2}\right)^{K-1}
\label{pitdsf}
\end{equation}
for $\omega\geqslant\omega_{-}(k)$, and zero otherwise. Within Luttinger-liquid theory, the dispersion is \emph{linear} near the umklapp point: $\omega_{-}(k)\simeq c|k-2 \pi n|$. By definition, $K\equiv \hbar\pi n/(m c)$ is the Luttinger parameter. For repulsive bosons, the value of parameter $K$ lies between $1$ (TG gas) and $+\infty$ (ideal Bose gas). The value of the Luttinger parameter in the strong-coupling regime
\begin{equation}
K=1 +4/\gamma + O(1/\gamma^2)
\label{Kstrong}
\end{equation}
is derived with the expression for the sound velocity obtained in Sec.~\ref{sec:linear}.
The coefficients in Eqs.~(\ref{den_den_t}) and (\ref{pitdsf}) are related by
\cite{PhD:astrakharchik04}
\begin{equation}
\widetilde{A}_1(K)=\frac{\Gamma^2(K)}{2\pi}\bigg(\frac{2K}{\pi}\bigg)^{2K}A_1(K)
\label{relA1k}
\end{equation}
with $\Gamma(K)$ being the gamma function. The coefficient $A_1(K)$ is non-universal,
and is explicitly known for the Lieb-Liniger model
\cite{2009_Kitanine_JSTAT_P04003,2010_Shashi_PREPRINT}.  We will make use of its value
in two limiting cases: $A_1(K)=\pi/4$ at $K=1$ and $A_1(K)\simeq
4^{1-3K}\exp(-2\gamma_{\mathrm{c}}K)\pi/\Gamma^{2}(K)$ for $K\gg 1$
\cite{astrakharchik04,PhD:astrakharchik04}, where $\gamma_{\mathrm{c}}=0.5772\ldots$ is the Euler constant.

By comparing the first-order expansion (\ref{DSFlinear}) in the vicinity of the umklapp point with Eq.~(\ref{pitdsf}) and using the expansion (\ref{Kstrong}), one can easily obtain \cite{cherny09} the model-dependent coefficient at large but \emph{finite} interactions when  $K-1\ll 1$
\begin{align}
\label{akseries} A_1(K) =\frac{\pi}{4}\Big[1 - \left(1+4\ln 2\right)(K-1)\Big]
 + O\left((K-1)^2\right).
\end{align}

Using the DSF (\ref{pitdsf}) predicted by Luttinger theory and equation (\ref{dragf1Dpoint}) for the drag force, we arrive at the expression for the drag force at small velocities $v\ll c$ \cite{astrakharchik04}
\begin{equation}
F_{\mathrm{v}}=\sqrt{\pi}\frac{\Gamma(K)}{\Gamma(K+1/2)}A_1(K)\frac{g^{2}_{\mathrm{i}}n^{2}}{\hbar
v} \left(2K\frac{v}{c}\right)^{2K}.
\label{DF_pit}
\end{equation}
In the TG regime, this formula yields the same values of the drag force as Eq.~(\ref{dragfTG}). Equation (\ref{DF_pit}) gives us the universal exponent of the power-law behaviour of the drag force at small velocities: $F_{\mathrm{v}}\sim v^{2K-1}$. The same result was obtained in Ref.~\cite{citro09}. While the non-universal coefficient $A_1(K)$ is now known for arbitrary strength of interactions \cite{2009_Kitanine_JSTAT_P04003,2010_Shashi_PREPRINT,2011_Shashi_PRB_84}, its actual expression is too unwieldy to be considered here. In Sec.~\ref{sec:effapprox} we prefer to consider another approach, which allows us to determine it approximately.

\subsection{The algebraic Bethe ansatz and ABACUS}
\label{sec:ABA}

\begin{figure}[t,b]
\includegraphics[width=.8\columnwidth]{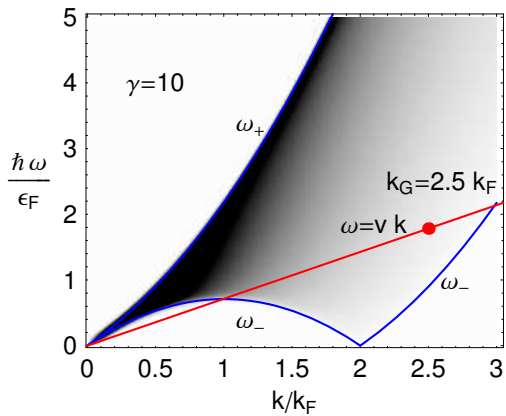}
\caption{\label{fig:omplmi} (Color online) Numerical values of the DSF (\ref{sqomega}) for the coupling parameter $\gamma = 10$ \cite{caux06}. The dimensionless value of the rescaled DSF $S(k,\omega)\varepsilon_{F}/N$ is shown in shades of gray between zero (white) and 1.0 (black). The upper and lower solid (blue) lines represent the dispersions $\omega_+(k)$ and $\omega_-(k)$, respectively, limiting the single ``particle-hole" excitations in the Lieb-Liniger model at $T=0$. The dispersions are obtained numerically by solving the system of integral equations \cite{lieb63:2}. The gray scale plot of the DSF demonstrates that the main contribution to the DSF comes from the single particle-hole excitations, lying inside the region $\omega_-(k)\leqslant\omega\leqslant\omega_+(k)$ (see also Fig.~\ref{fig:schem}). Only one point at $k=k_G$, shown in full (red) circle, contributes to the integral when the perturber is a shallow cosine potential with a reciprocal vector $k_G$. Reproduced from Ref.~\cite{cherny09a}, Copyright \copyright 2009 The American Physical Society}
\end{figure}

The exact integrability of the Lieb-Liniger model now permits the direct numerical calculation of dynamical correlation functions such as the DSF \cite{caux06} for systems with finite numbers of particles using the ABACUS algorithm \cite{2009_Caux_JMP_50}.  The strategy consists in using the Lehmann representation (\ref{sqomega0}).  The eigenstates themselves are constructed by the Bethe Ansatz.  The state norms \cite{gaudin83book,1982_Korepin_CMP_86} and matrix element of the density operator \cite{1989_Slavnov_TMP_79,1990_Slavnov_TMP_82} being exactly known, the sum over intermediate states can be taken starting from the dominant few-particle states and let run until a satisfactory saturation of sum rules is achieved.  The results for the DSF take the form illustrated in Fig. \ref{fig:omplmi}. For generic values of $\gamma$, most of the signal is concentrated between the dispersions $\omega_+ (k)$ and $\omega_-(k)$ of single particle/hole (Type 1/2 in Lieb's notation) excitations.  The signal is by construction identically zero below $\omega_-(k)$ (since no eigenstates are found there);  above $\omega_+(k)$, the signal is carried by multiple particle-hole excitations and is very weak in view of the small matrix elements of the density operator between these states and the ground state.

We use data for $N = 150$ particles ($\gamma = 5$ and $20$), $N = 200$ ($\gamma = 1$) and $N = 300$ ($\gamma = 0.25$).  The $f$-sum rule saturations at $k = 2k_F$ were $99.64 \%$ ($\gamma = 20$), $97.81 \%$ ($\gamma = 5)$, $99.06 \%$ ($\gamma = 1$) and $99.08 \%$ ($\gamma = 0.25$), the saturation getting better at smaller momenta. The drag force is computed from the numerical DSF data in the following way. Since the DSF in finite size is given by discrete but densely distributed $\delta$-function peaks, we consider the integral
\begin{align}
&F^I (v) \equiv \frac{g^2_\mathrm{i}}{L}\int_0^\infty \d q \int_0^{qv} d\omega S (q, \omega)\nonumber\\
&= \frac{g^2_\mathrm{i}}{\hbar L}\int_0^\infty \d q \sum_{n} |\langle0|\delta\hat{\rho}_{q} |n\rangle|^2 \Theta(\hbar qv - E_n + E_0),
\end{align}
whose derivative with respect to $v$ simply gives the drag force (\ref{dragf1Dpoint}). Here $\Theta$ is the Heaviside step function. The integral in $\omega$ conveniently gets rid of all energy $\delta$-functions in the expression for the DSF, and this integrated quantity is readily computed without need for smoothing using the raw ABACUS data for the DSF.  The derivative with respect to $v$ can then be taken numerically by fitting an interpolating polynomial to the data points for $F^I (v)$ in the vicinity of the velocity for which the drag force needs to be calculated.  The resulting data for the drag force are illustrated in Fig.~\ref{fig:drag}a.

The advantage of this method is that the reliability of the results is more or less independent of the value of the interaction parameter, in the sense that states and matrix elements can be individually constructed irrespective of what $\gamma$ is.  The remaining issues are the distribution of correlation weight among the excitations, and how this affects speed of convergence.  The limit of very small $\gamma$ is the easiest to treat, since only very few states in the vicinity of the Type 1 mode are of importance.  The DSF then becomes almost a $\delta-$function in the Type 1 dispersion relation, and the drag force tends to the step function as expected in this limit.  The opposite case of infinite $\gamma$ is also straightforward, since then only single particle-hole excitations have non-negligible matrix elements, the drag force becomes a constant between Type 1 and 2 dispersions, and the drag force acquires a constant slope. For interaction values between these two extremes, many eigenstates must be summed over for good convergence, and the DSF takes on a nontrivial lineshape making the drag force take the positive-curvature shapes in Fig.~\ref{fig:drag}a.

\subsection{An effective approximation for the dynamic structure factor and drag force}
\label{sec:effapprox}

In Ref.~\cite{cherny08,cherny09}, an interpolating expression was suggested for the DSF:
\begin{equation}
\label{dsfapp1} S(k,\omega)=C
\frac{(\omega^{\alpha}-\omega_{-}^{\alpha})^{\mu_{-}}}
{(\omega_{+}^{\alpha}-\omega^{\alpha})^{\mu_{+}}}
\end{equation}
for $\omega_{-}(k)\leqslant\omega\leqslant\omega_{+}(k)$, and $S(k,\omega)=0$ otherwise. Here, $\mu_{+}(k)$ and $\mu_{-}(k)$ are the exact exponents \cite{imambekov08}\footnote{Note that the field theory predictions of \cite{imambekov08} actually include a singularity also for $\omega > \omega_+(k)$, with a universal shoulder ratio. We neglect this here since it gives only a small correction to the results.}
\begin{equation}
S(k,\omega)\sim \big|\omega-\omega_{\pm}(k)\big|^{\mp \mu_{\pm}(k)}
\label{glazexp}
\end{equation}
at the borders of the spectrum $\omega_{+}(k)$ and $\omega_{-}(k)$. We also put by definition $\alpha\equiv 1+1/\sqrt{K}$. The most general way of obtaining $\omega_{\pm}(k)$, $\mu_{\pm}(k)$, and $K$ is to solve numerically the corresponding integral equations of Refs.~\cite{lieb63:2} and \cite{imambekov08}, respectively.

It follows from energy and momentum conservation that $S(k,\omega)$ is exactly equal to zero below $\omega_{-}(k)$ for $0\leqslant k \leqslant 2 \pi n$. In the other regions of $\omega > \omega_{+}$ and  $\omega < \omega_{-}$ (for $k > 2 \pi n$), possible contributions can arise due to coupling to multi-particle excitations \cite{lieb63:2}. However, these contributions are known to vanish in the Tonks-Girardeau ($\gamma \to \infty$) and Bogoliubov ($\gamma \to 0$) limits and are found to be very small numerically for finite interactions \cite{caux06}.

\begin{figure}[tbh]
\centerline{\includegraphics[width=0.75\columnwidth]{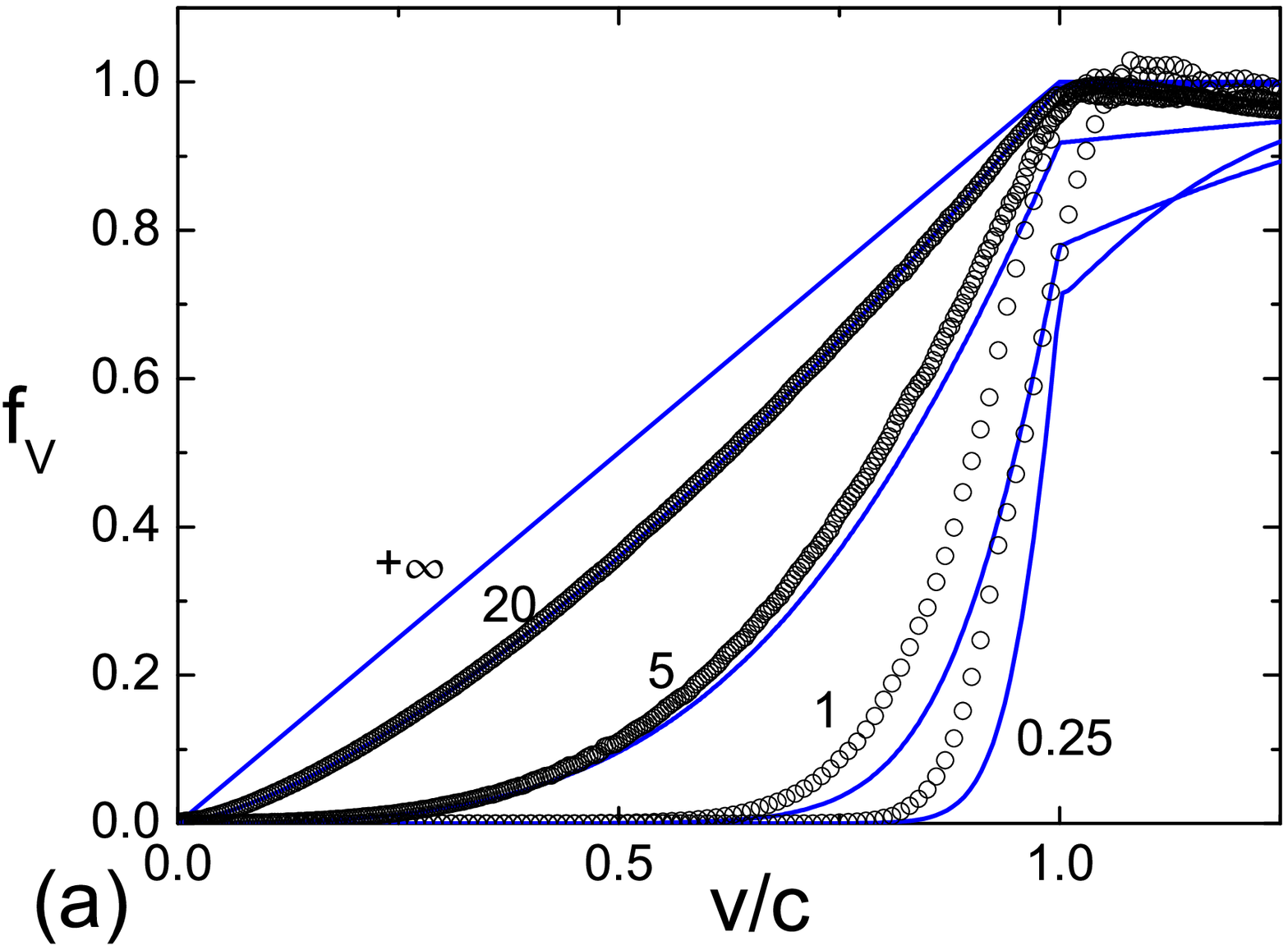}}
\centerline{\includegraphics[width=0.775\columnwidth]{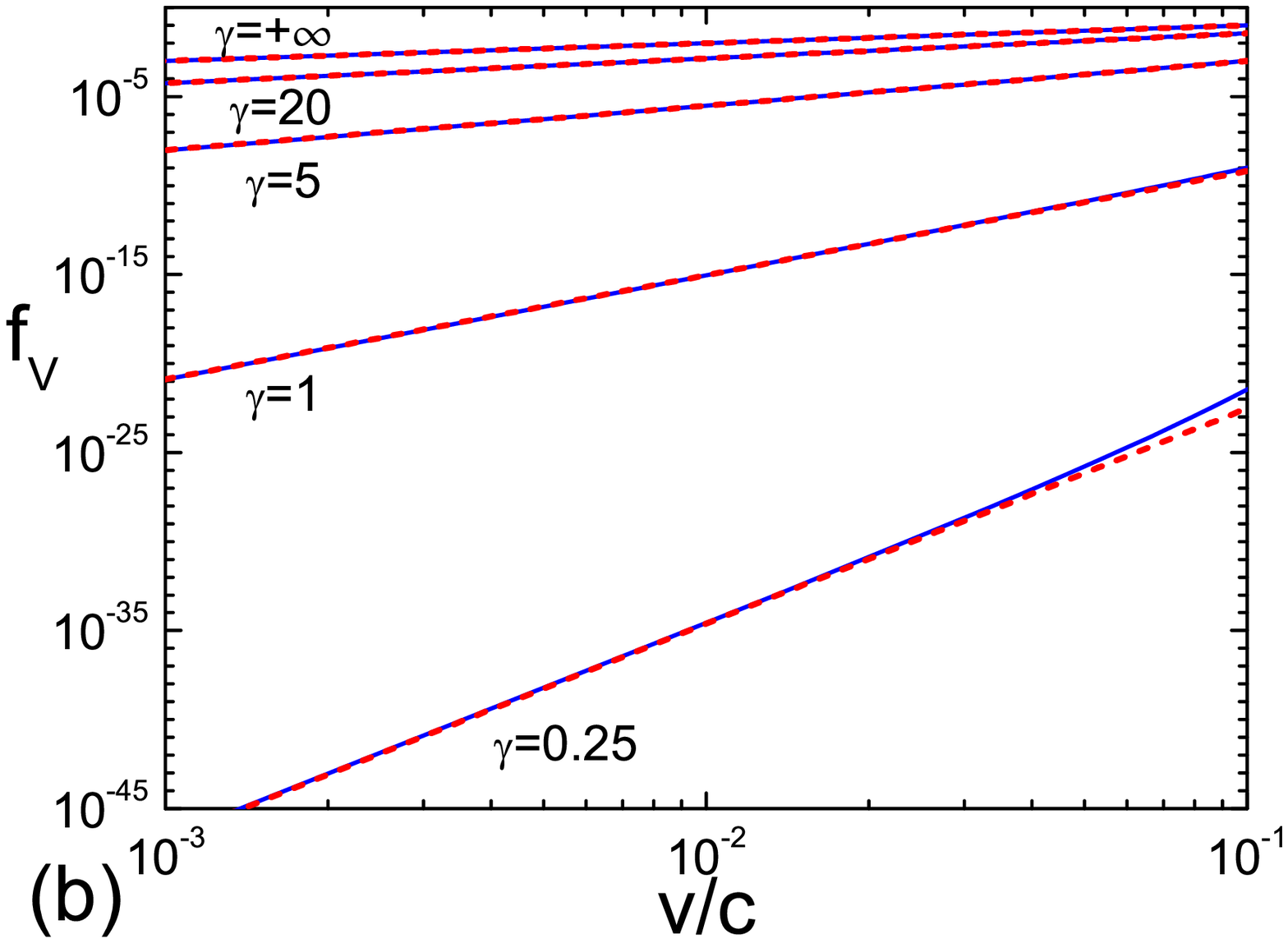}}
\caption{\label{fig:drag} (Color online) The dimensionless drag force versus the velocity (relative to the sound velocity) of the impurity at various values of the coupling parameter $\gamma$. (a) The solid (blue) lines represent the force obtained with Eqs.~(\ref{dragf1Dpoint}) and (\ref{dsfapp1}), open circles are the numerical data obtained using ABACUS \cite{caux06}. (b) The values of the dimensional drag force obtained from the interpolation formula (\ref{dsfapp1}) [solid (blue) lines] are compared with that of the analytical formula (\ref{DF_universal1}) for small velocities [dashed (red) lines].}
\end{figure}

The exponents $\mu_{\pm}$ are non-negative \footnote{We slightly change the notations: our $\omega_{\pm}$ and $\pm\mu_{\pm}$ correspond to $\omega_{1,2}$ and $\mu_{1,2}$ in Ref.~\cite{imambekov08}, respectively. We also denote the density of particles $n$ and the Fermi wavevector for quasiparticles $q_0$ instead of $D$ and $q$ used in Refs.~\cite{korepin93:book,imambekov08}, respectively.}.
As a consequence, the DSF diverges at the upper branch $\omega_{+}$. At the lower branch $\omega_{-}$, the DSF shows a continuous transition to zero for any finite value of $\gamma$ except for the specific point $\gamma = +\infty$ (or $K=1$) of the Tonks-Girardeau gas, where the DSF remains finite but has a discontinuous transition to zero at both boundaries $\omega_{-}$ and $\omega_{+}$.

The normalization constant $C$ depends on momentum but not on frequency and is determined from the $f$-sum rule (\ref{fsumrule}). The expression (\ref{dsfapp1}) is applicable for all ranges of the parameters $k$, $\omega$, and $\gamma$ with increasing accuracy at large $\gamma$ \cite{cherny09}.

The parameter $\alpha$ is needed to reconcile the limiting value of the exponent $\mu_{-}(2\pi n)=2\sqrt{K}(\sqrt{K}-1)$ in the vicinity of the umklapp point \cite{imambekov08} and the Luttinger theory predictions, given by Eq.~(\ref{pitdsf}). Now one can see from (\ref{dsfapp1}) that
\begin{equation}
\label{glazdsfexp} S(k,\omega)\sim \left\{\begin{array}{ll}
\omega^{2(K-1)},& k=2\pi n,\\
(\omega-\omega_{-})^{\mu_{-}(k)},& k\not=2\pi n.
\end{array}\right.
\end{equation}
Thus, the suggested formula (\ref{dsfapp1}) is consistent with both the Luttinger liquid behavior at the umklapp point and Imambekov's and Glazman's power-law behavior in its vicinity, as it should be. A more detailed discussion can be found in Ref.~\cite{cherny09}. Similar approximations for the DSF of 1D Bose gas, confined in a harmonic trap, are considered in Ref.~\cite{golovach09}.

The drag force can now be calculated by means of Eqs.~(\ref{dragf1Dpoint}) and (\ref{dsfapp1}) for arbitrary strength of interactions and arbitrary velocities. The results are shown in Fig.~\ref{fig:drag}a.

For the important question whether persistent currents may exist at all, the small velocity regime is most relevant, which is dominated by transitions near the first supercurrent state (umklapp point at $\omega=0$ and $k = 2\kf$). The drag force in this regime has a power-law dependence on the velocity $F_\mathrm{v}\sim v^{2K-1}$ for $v\ll c$, as first found by Astrakharchik and Pitaevskii \cite{astrakharchik04}. From Eqs.~(\ref{dragf1Dpoint}) and (\ref{dsfapp1}) we obtain
\begin{align}
&f_\mathrm{v}\equiv\frac{F_\mathrm{v}\pi\ef}{g_\mathrm{i}^2 \kf^3} \simeq 2K
\bigg(\frac{v}{v_\mathrm{F}}\bigg)^{2K-1}\bigg(\frac{4\ef}{\hbar\omega_{+}(2\kf)}\bigg)^{2K}\nonumber \\
&\times\!\!\frac{\Gamma\big(1+\frac{2 K}{\alpha}-\mu_{+}(2\kf)\big)}{\Gamma\big(\frac{2 K}{\alpha}\big)\Gamma\big(1-\mu_{+}(2\kf)\big)}
 \frac{\Gamma\big(1+\mu_{-}(2\kf)\big)\Gamma\big(1+\frac{1}{\alpha}\big)}{\Gamma\big(1+\mu_{-}(2\kf)+\frac{1}{\alpha}\big)},
\label{DF_universal1}
\end{align}
where $\Gamma(x)$ is Euler's Gamma-function, and $\mu_{-}(2\kf) =2\sqrt{K}(\sqrt{K}-1)$~\cite{imambekov08}. This formula is valid for {\it arbitrary} coupling constant and works even in the Bogoliubov regime at $\gamma\ll 1$. In practice, Eq.~(\ref{DF_universal1}) works well up to $v\lesssim 0.1 c$, see Fig.~\ref{fig:drag}b. The strong suppression of the drag force in the Bogoliubov regime appears because of the large exponent value $2K -1\gg 1$ and the large argument $2K/\alpha\gg 1$ of the gamma function in the denominator. Certainly, drag force values of order $10^{-40}$ lie beyond realizable experiments and as a matter of fact, tell us about superfluidity in this regime.

\subsection{Drag force from the phase slip transitions}
\label{sec:slip}

The quantum number $I$, describing the supercurrent with energy (\ref{EI}), is nothing else but the phase winding number of the original Bose field $\Psi=\sqrt{n}e^{i \Phi}$ \cite{haldane81}. The transitions between the supercurrent states are the phase slip transitions, changing the phase winding number. One can write down explicitly the low-energy Hamiltonian \cite{kashurnikov96} that consists of the supercurrent states and phonon excitations, interacting due to a static impurity. The interaction $V_\mathrm{i}(x)$ between the impurity and bosons
can be rewritten in terms of the supercurrent states and the phonons. One can then directly derive the transition probability per unit time between the supercurrent states $I$ and $I'$ from Fermi's golden rule \cite{kagan00}. The result reads at zero temperature [see Eq.~(8) of Ref.~\cite{kagan00}]
\begin{align}
W_{II'}= \frac{|g g_\mathrm{i} n|^2}{\hbar\,\Gamma [1+\alpha] \Gamma
[\frac{1}{2}+\alpha]} \frac{4\pi^{3/2}\alpha}{\xi_{II'}} \bigg(\frac{\xi_{II'}}{\widetilde{\gamma}
\epsilon_0}\bigg)^{2\alpha}, \label{wii}
\end{align}
where $\alpha\equiv(I-I')^2 K$ and $\xi_{II'}\equiv \varepsilon_{I}-\varepsilon_{I'}$. The transition probability (\ref{wii}) is determined up to an overall factor, because the dimensionless parameters $g$ and $\widetilde{\gamma}$, which are of the order of 1, cannot be defined exactly from the long-range effective Hamiltonian. The energy $\epsilon_0$ is a characteristic high-energy cutoff for the phonon spectrum.

For the $I$th supercurrent decay at zero temperature due to the static impurity, the dominant contribution comes from the transition to the $(I-1)$th state, and one can write for the energy loss per unit time
\begin{align}
\dot{E}=W_{I,I-1}\xi_{I,I-1}= \frac{|g g_\mathrm{i} n|^2\,4\pi^{3/2}K}{\hbar\,\Gamma [1+K] \Gamma
[\frac{1}{2}+K]} \bigg(\frac{\xi_{I,I-1}}{\widetilde{\gamma} \epsilon_0}\bigg)^{2K}.
\label{elossI}
\end{align}
For large winding number $I$, we have the relation $\xi_{I,I-1}\simeq 2\pi n \hbar v_I$, resulted from Eq.~(\ref{EI}). By putting the energy cutoff $\epsilon_0$ to be equal to $2\pi n \hbar c$ and remembering the definition of the drag force $|\dot{E}|=F_\mathrm{v} v$, we obtain from Eq.~(\ref{elossI})
\begin{equation}
F_v=|g g_\mathrm{i} n|^2\frac{4\pi^{3/2}}{\hbar\,\Gamma[K]\,\Gamma
[\frac{1}{2}+K]} \frac{1}{v} \bigg(\frac{v}{\widetilde{\gamma} c}\bigg)^{2K}.
\label{dfI}
\end{equation}
Thus, we come to the dependence $F_v\sim v^{2K-1}$ at small velocities, in agreement with the previous results (\ref{DF_pit}) and (\ref{DF_universal1}).

\subsection{Direct calculation for the Tonks Girardeau gas beyond linear response theory}
\label{sec:beyond_linear}

\begin{figure}[tbhp]
\includegraphics[width=.8\columnwidth]{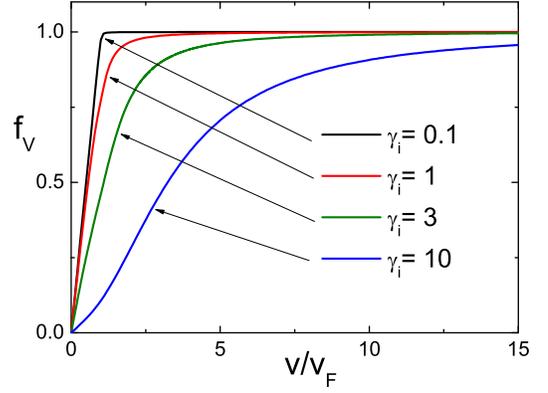}
\caption{\label{fig:Iv_Pi_gamma_ksi} The diagram shows the dimensionless drag force (\ref{dragf1Dpoint}) for the TG gas at various values of the {\it impurity} coupling parameter $\gamma_\mathrm{i}\equiv g_\mathrm{i} m/(\hbar^2 n)$. The results are obtained \emph{beyond} the linear response approximation.  Note that the absolute value of the drag force $F_\mathrm{v}$ is proportional to $\gamma_\mathrm{i}^2$. }
\end{figure}

In this section we consider the TG gas, or the ideal gas of spin-polarized fermions. In order to find the drag force, we choose a frame of reference, in which the massive impurity is at rest. Moving particles are then scattered by the impurity potential $V_\mathrm{i}(x) =g_\mathrm{i}\delta(x)$. For an {\it ideal} Fermi gas the scattering process for each particle is {\it independent} of the scattering of the other particles. The reflection coefficient for a particle with wave vector $k$ is easily determined from the one-body Schr\"odinger equation: $r(k)=m^2g_\mathrm{i}^2/(m^2g_\mathrm{i}^2 +\hbar^4k^2)$. If $\d N_{k}$ is the number of particles with wave vectors lying between $k$ and $k+\d k$, then these particles transfer momentum to the impurity per unit time $\d F=2\hbar^2 n r(k)k^2 \d N_{k}/m$. Since in the chosen frame of reference, the momentum distribution is the Fermi distribution shifted by $m v$, we can write down the explicit expression of the drag force
\begin{equation}
F_\mathrm{v}=\frac{2\hbar^2}{m}\int \frac{\d k}{2\pi}n(k-mv/\hbar) \,\mathrm{sign}(k)
r(k)k^2, \label{DF_non_lin}
\end{equation}
where $n(k)=1$ for $-\pi n\leqslant k\leqslant \pi n$, and zero otherwise. The integral
is readily taken
\begin{align}
&F_\mathrm{v}=\frac{2m n g_\mathrm{i}^2}{\hbar^2}\times\nonumber\\
&\left\{ \begin{array}{ll}
\xi +\frac{\gamma_\mathrm{i}}{2\pi}\left(\arctan\frac{\pi(1-\xi)}{\gamma_\mathrm{i}}-\arctan\frac{\pi(1+\xi)}{\gamma_\mathrm{i}}\right),
& 0\leqslant\xi\leqslant 1 ,\\[2ex]
1 +\frac{\gamma_\mathrm{i}}{2\pi}\left(\arctan\frac{\pi(\xi-1)}{\gamma_\mathrm{i}}-\arctan\frac{\pi(1+\xi)}{\gamma_\mathrm{i}}\right),
& \xi\geqslant 1.
\end{array}\right.
\label{DF_non_lin1}
\end{align}
Here we denote $\gamma_\mathrm{i}\equiv g_\mathrm{i} m/(\hbar^2 n)$, and $\xi\equiv v/v_\mathrm{F}$. The results are shown in Fig.~\ref{fig:Iv_Pi_gamma_ksi}. As discussed in Sec.~\ref{sec:largev}, the limiting value of the drag force at large velocities coincides with that obtained within linear response theory. At small impurity coupling $\gamma_\mathrm{i}\ll 1$, Eq.~(\ref{DF_non_lin1}) reproduces the linear response formula (\ref{dragfTG}), see Fig.~\ref{fig:DF_var_approx}a.

\begin{figure}[tbhp]
\center{\hspace*{3em}\includegraphics[width=.5\columnwidth,clip=true]{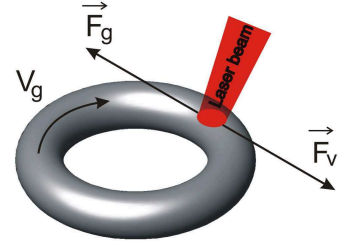}}\\
\caption{\label{fig:exper_scheme} An experimental scheme to observe current decays. An artificial ``impurity" is created by optical methods. It can be switched on adiabatically or abruptly.}
\end{figure}

\section{Consquence of drag force: velocity damping}
\label{sec:velocity_damping}

\begin{figure}[tbhp]
\center{\includegraphics[width=.8\columnwidth]{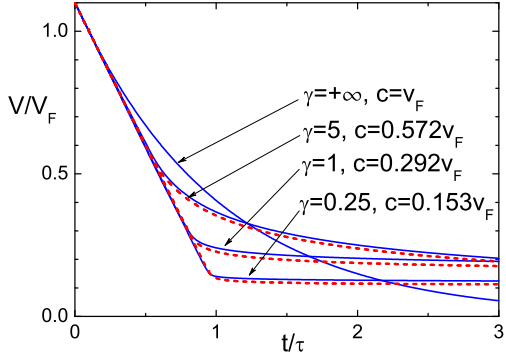}}
\caption{\label{fig:exper_damping} (Color online) Decay of the ring current velocity of 1D bosons from the initial velocity of $1.1 \vf$ at $t=0$. The solid (blue) and dashed (red) lines represent the results obtained with the approximate formula and ABACUS, respectively. The time scale is ${\tau}={N \pi \hbar^3}/{(2m g_{\mathrm{i}}^{2})}$. Reproduced from Ref.~\cite{cherny09a}, Copyright \copyright 2009 The American Physical Society}
\end{figure}

In the presence of an obstacle as, e.g., shown in Fig.~\ref{fig:exper_scheme}, a ring current can decay due to successive transitions to supercurrent states with smaller momentum. Starting in one of the local minima of the excitation spectrum as seen in Fig.~\ref{fig:schem}, the kinetic energy of the center-of-mass translation will be transformed into elementary excitations above a lower supercurrent state while still conserving the total energy. The elementary excitations are quasiparticle-quasihole excitations in the Bethe-ansatz wave function \cite{lieb63:2}. Both in energy and character most of these excitation lie between the phonon-like $\omega_+$ branch and the $\omega_-$ branch, which is related to dark solitons (see Fig.~\ref{fig:omplmi}).
Assuming that these excitations have little effect on successive transitions, we estimate \cite{cherny09a} the decay of the center-of-mass velocity $v$ by the classical equation $Nm \dot{v} = -F_\mathrm{v}(v)$, where $F_\mathrm{v}$ is given by Eqs.~(\ref{dragf1Dpoint}) and (\ref{dsfapp1}). This equation was integrated numerically and the result is shown in Fig.~\ref{fig:exper_damping}. At the initial supersonic velocity, where the drag force is saturated (see Fig.~\ref{fig:drag}a) the supercurrent experiences constant deceleration. For $v\lesssim c$ the drag force decreases and consequently the deceleration slows down. For the TG gas we find an analytical solution for exponential decay  $v(t)=v_0\exp(-t/\tau)$  for  $v_0\leqslant v_\mathrm{F}$. In the weakly-interacting regime, the decay may be slow compared to experimental time scales.

\section{Drag force in extended potentials}
\subsection{1D bosons in a moving shallow lattice}
\label{sec:shallow_lattice}
\begin{figure}[tbhp]
\centerline{\includegraphics[width=.75\columnwidth,clip=true]{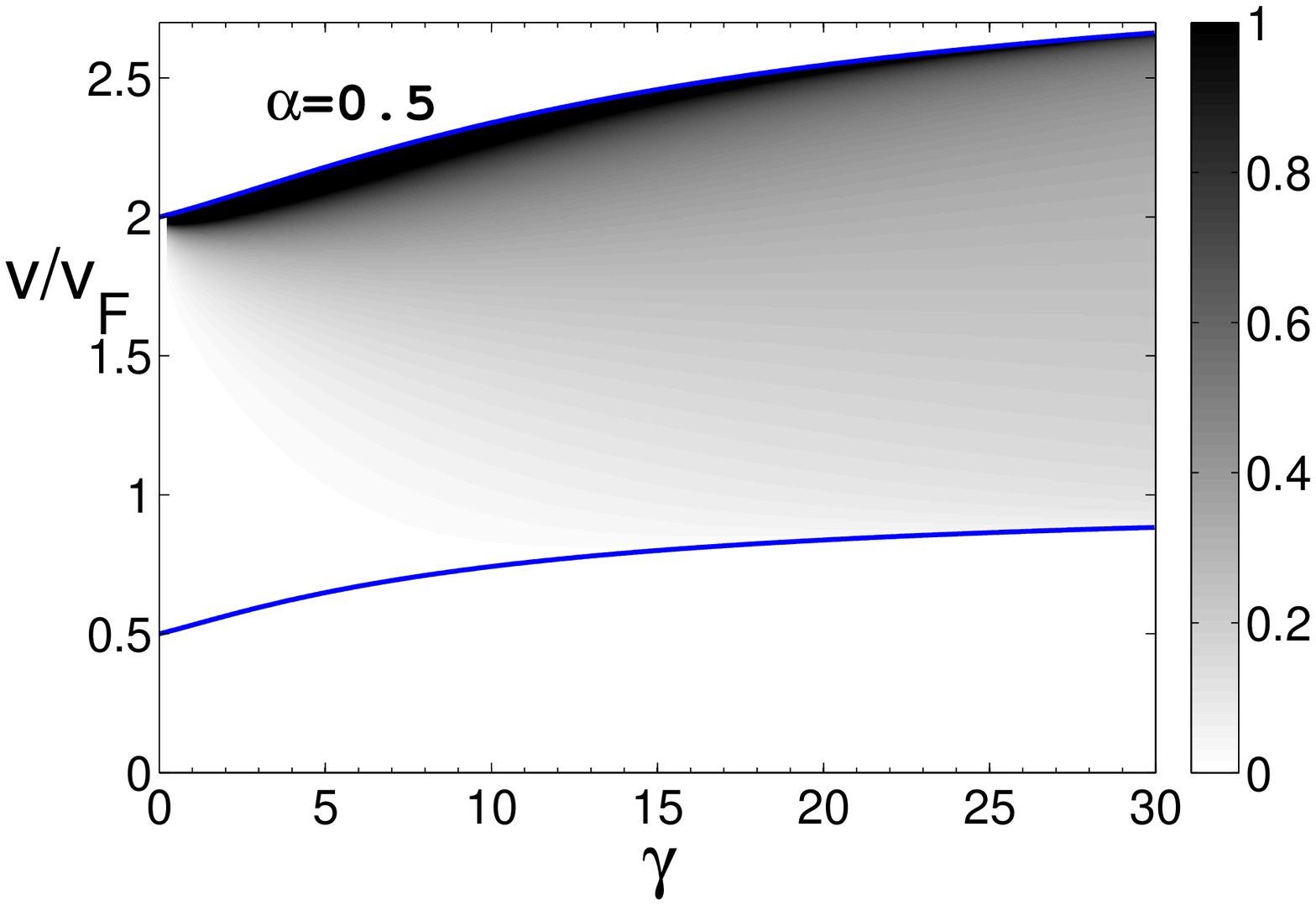}}
\centerline{\includegraphics[width=.75\columnwidth,clip=true]{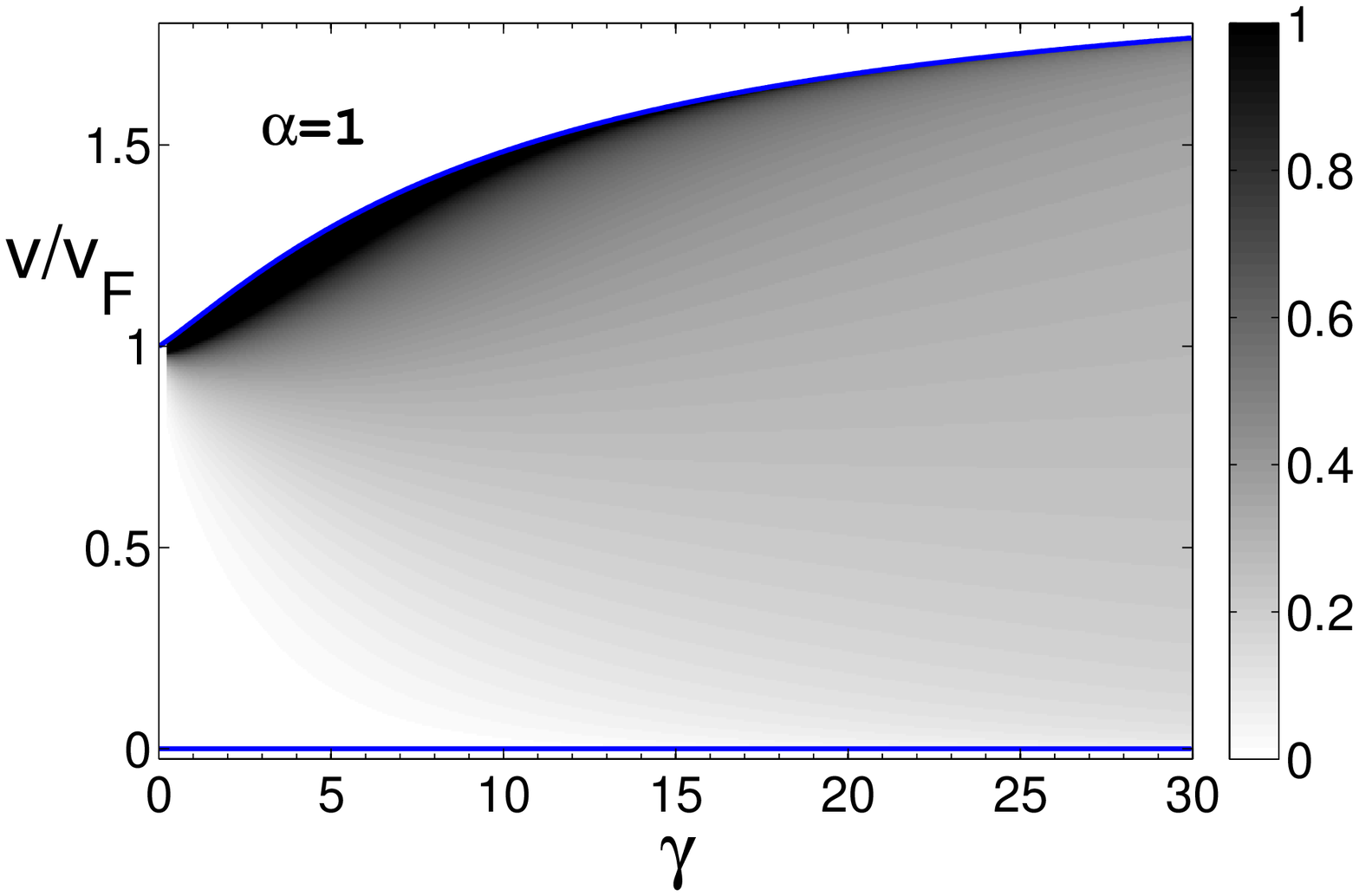}}
\centerline{\includegraphics[width=.75\columnwidth,clip=true]{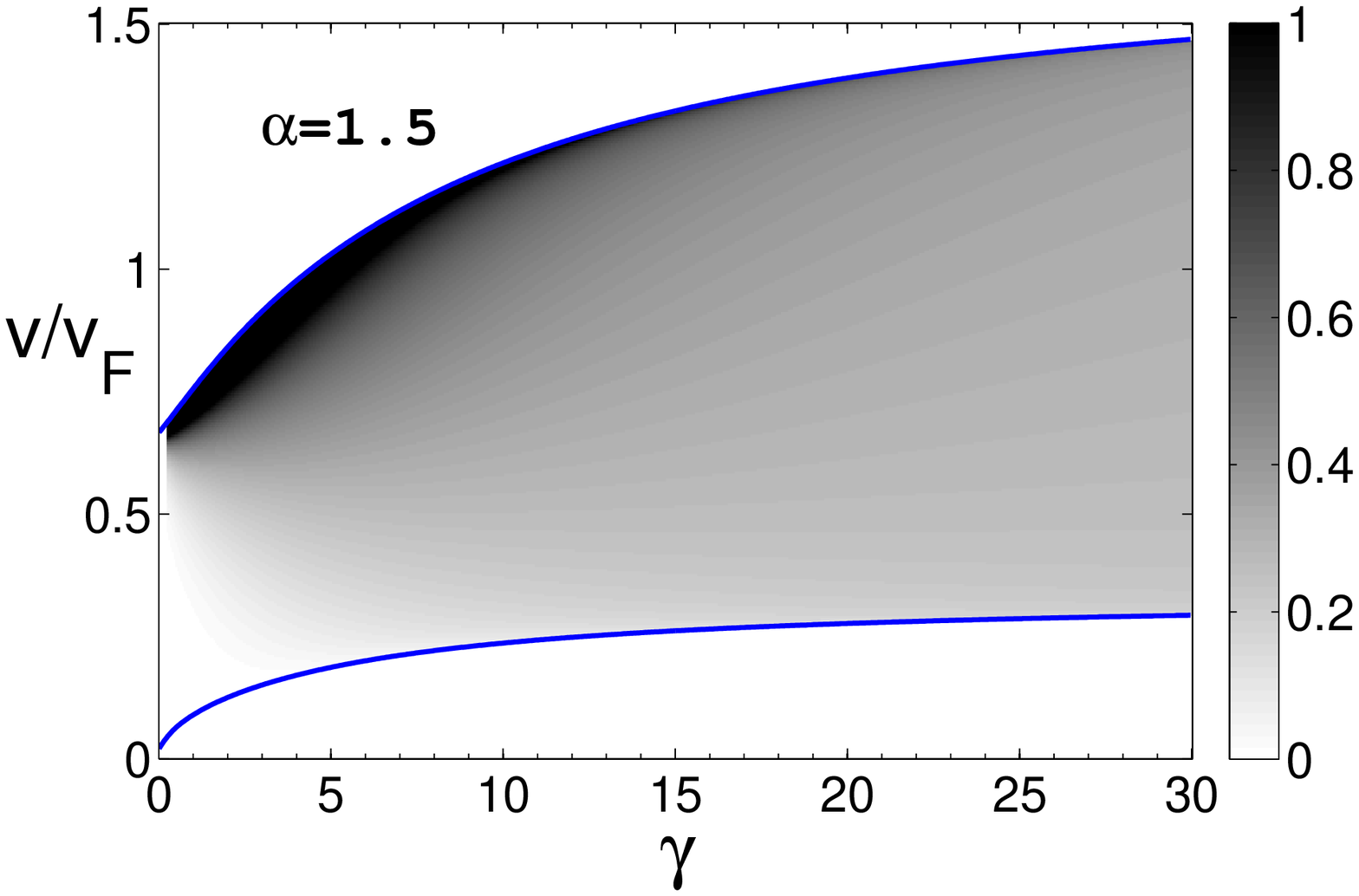}}
\caption{\label{fig:vgam} (Color online) Zero temperature phase diagram for superfluid-isolator transition of the Bose gas in a moving shallow lattice: dimensionless drag force $F_\mathrm{v}\hbar^2/(g^2_\mathrm{L}mL)$ versus the lattice velocity (in units $v_\mathrm{F}$) and the interaction strength $\gamma$ at various values of the filling factor $\alpha$. The dimensionless values are  represented in shades of gray between zero (white) and 1.0 (black). The solid (blue) lines correspond to the DSF borders $\omega_+(k)$ and $\omega_-(k)$, respectively.}
\end{figure}

Equation (\ref{dragf1D}) can be verified experimentally for different types of obstacles:  for $V_\mathrm{i}(x)=g_\mathrm{i}\delta(x)$ all the points at the transition line contribute to the drag force, while for the periodic potential with the spatial period $a$ only a set of discrete points in the $\omega$-$k$ plane do \cite{cherny09a,cherny10a}.  This is simply due to the fact that a periodic potential has only a discrete set of Fourier components with momenta $j k_G$ with $j$ being integer and  $k_G\equiv 2 \pi/a$ being the reciprocal lattice vector. For instance, we have two nonzero components $k_G$ and $-k_G$ for optical lattice potential $V_\mathrm{i}(x)= g_\mathrm{L}\cos(2\pi x/a)$
\begin{equation}
|\widetilde{V}_\mathrm{i}(q)|^2/L=\pi
g^2_\mathrm{L}\big(\delta(q-k_G)+\delta(q+k_G)\big)/2 \label{V2kG}
\end{equation}
in the thermodynamic limit ($n=\mathrm{const}$, $L\to\infty$). The filling factor of the lattice, that is, the number of particles per site, is given by
$\alpha=2\pi n/k_G$, because the total number of lattice periods equals $L k_G/(2\pi)$. Substituting Eq.~(\ref{V2kG}) into the general expression for the drag force (\ref{dragf1D}), we obtain at zero temperature
\begin{equation}
F_{\mathrm{v}} = \pi g_{\mathrm{L}}^{2} k_G S(k_G,k_G v)/2.
\label{dfkG}
\end{equation}
The values of the obtained drag force can easily be estimated with the $\omega$-$k$ diagram for the DSF, see Fig.~\ref{fig:omplmi}. Note that the drag force is now proportional to the total number of particles, by contrast to the point-like impurity case, described by Eq.~(\ref{dragf1Dpoint}). This is because the lattice potential is non-local.

Let us recall that the model considered in this paper assumes periodic boundary conditions. However, according to the general principles of statistical mechanics, the boundary conditions do not play a role in the thermodynamic limit. Hence, the obtained formula (\ref{dfkG}) can be exploited at sufficiently large number of particles even in the case of a cigar-shaped quasi-1D gas of bosons. This equation gives us the momentum transfer per unit time from a moving shallow lattice, which can be measured experimentally \cite{fallani04,fertig05,mun07}.

In the case of Bogoliubov and TG regimes, the drag force admits analytic solutions. As discussed in Sec.~\ref{sec:bogregime}, at small $\gamma$ the upper dispersion curve $\omega_{+}(k)$ is described well by the Bogoliubov relation (\ref{bogdisp}), and the behavior of the DSF simulates the $\delta$-function spike in accordance with Eq.~(\ref{dsfbog}). We derive from Eqs.~(\ref{dsfbog}) and (\ref{dfkG})
\begin{equation}
F_\mathrm{v}=\frac{g_\mathrm{L}^{2}m L}{\hbar^2}\frac{1}{2\alpha\xi}
\delta\bigg(\xi-\sqrt{\frac{1}{\alpha^2}+\frac{\gamma}{\pi^2}}\bigg),
\label{dfkGBog}
\end{equation}
where by definition $\xi\equiv v/v_\mathrm{F}$.

In the TG regime, discussed in Sec.~\ref{sec:TG}, the DSF is given by Eq.~(\ref{DSFTG}) with the limiting dispersions $\omega_\pm$ being known analytically.
Then Eqs.~(\ref{DSFTG}) and (\ref{dfkG}) yield
\begin{equation}
F_\mathrm{v}=g_\mathrm{L}^{2}m L/(4\hbar^2)[\Theta(\xi-\xi_+)-\Theta(\xi-\xi_-)],
\label{dfkGTG}
\end{equation}
where we put by definition $\xi_\pm\equiv|1\pm 1/\alpha|$.

\begin{figure}[tb]
\centerline{\includegraphics[width=.8\columnwidth,clip=true]{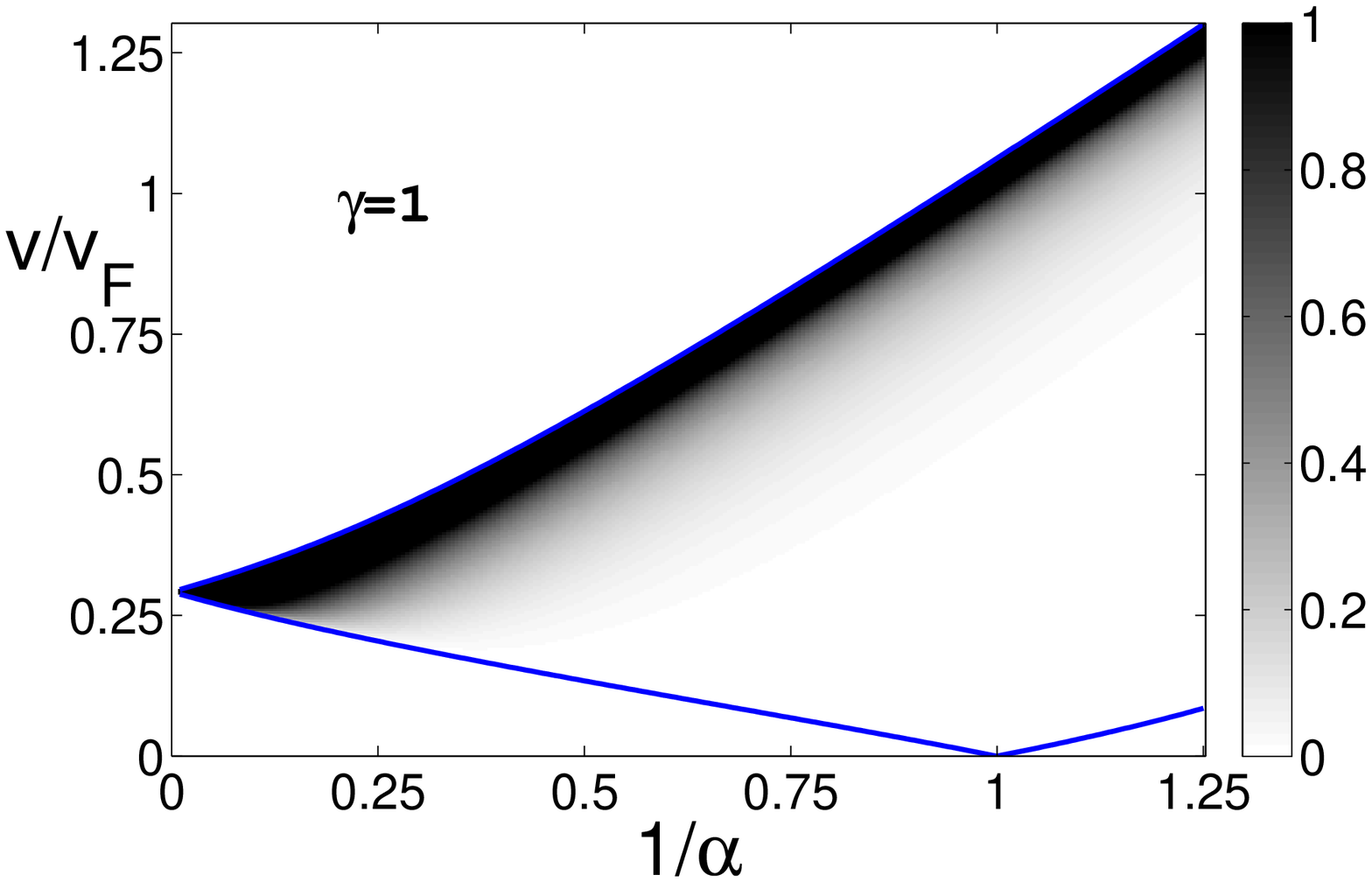}}
\centerline{\includegraphics[width=.8\columnwidth,clip=true]{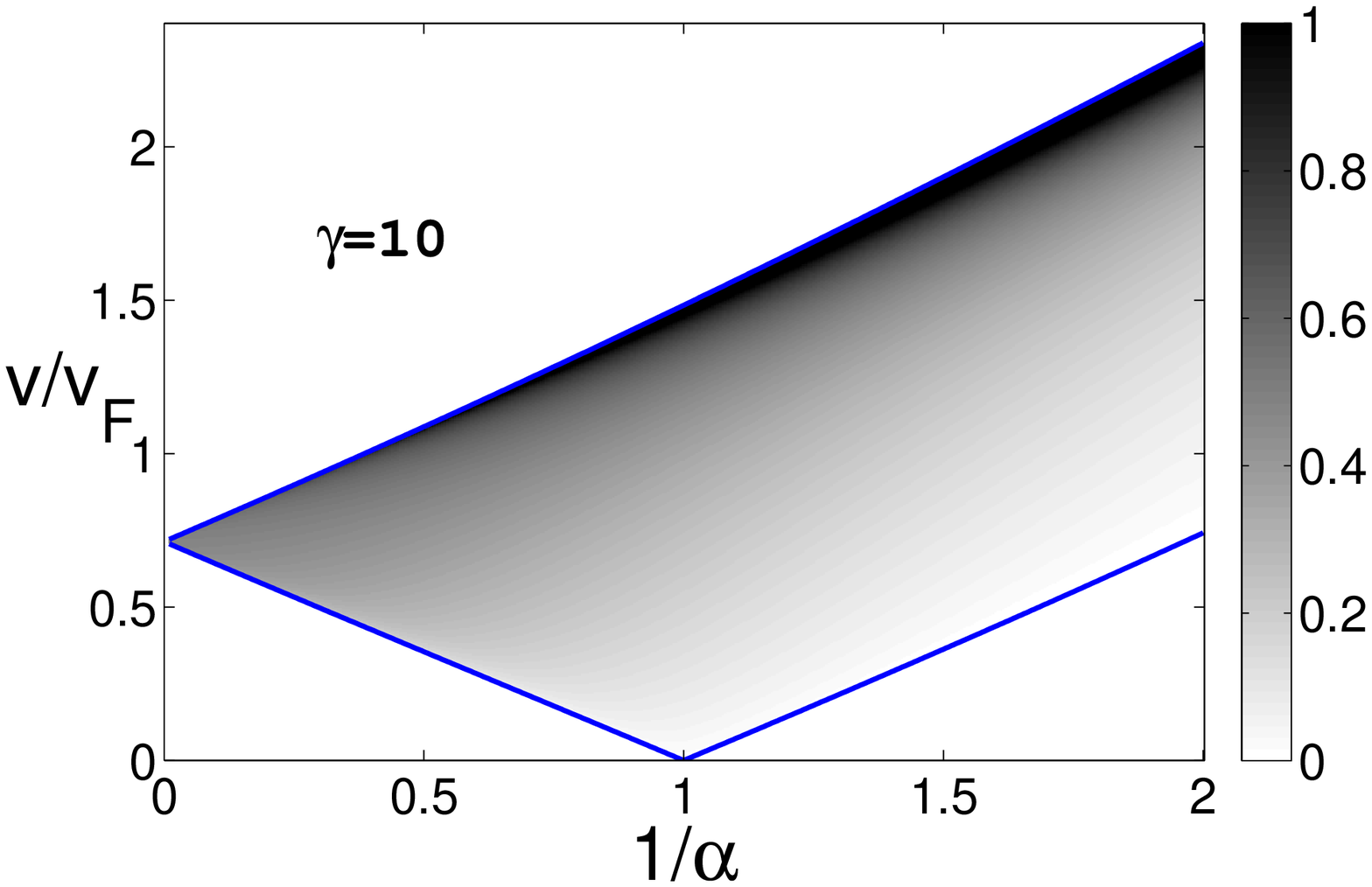}}
\caption{\label{fig:valph} (Color online) The same diagram as shown in Fig.~\ref{fig:vgam}, but here the drag force is represented as a function of velocity and inverse filling factor.}
\end{figure}

The values of the drag force (\ref{dfkG}) obtained from the interpolating expression for DSF (\ref{dsfapp1}) is shown in Figs.~\ref{fig:vgam} and \ref{fig:valph}. The filling factor  $\alpha=1$ ($k_G=2\kf$) corresponds to the Mott insulator state in a deep lattice. As can be seen from Fig.~\ref{fig:omplmi}, at this value of the reciprocal vector the DSF is almost independent of $\omega$ when $\gamma \gg1$, and the drag force takes non-zero values for arbitrary $v\leqslant \omega_+(k_G)/k_G$, see Fig.~\ref{fig:vgam}.

By contrast, at small $\gamma$ its non-zero values practically localize in the vicinity of $v=\omega_+(k_G)/k_G$, as shown in Fig.~\ref{fig:valph}. Then superfluidity breaks down when the point $\omega=v k_G$ and $k=k_G$ lies exactly on the Bogoliubov dispersion curve, see Fig.~\ref{fig:omplmi}. Taking into account that the Bogoliubov dispersion is very close to the free particle one, we obtain for the break point $k_G v= \hbar k_G^2/(2m)$. One can see that this point coincides with the point of dynamical instability for bosons in the cosine shallow  lattice. Indeed, the dynamical instability appears at a value of lattice velocity $v$ corresponding to a negative curvature of the Bloch dispersion curve at the momentum $mv/\hbar$. This is because in the reference frame where the lattice is at rest, particles' momenta are shifted by the value $mv/\hbar$, and the effective mass, given by the second derivative of the curve in this point, becomes negative. This implies that the effective kinetic energy of particles and the interaction energy have different signs, which immediately leads to the instability of the repulsive Bose gas at sufficiently large numbers of particles. For a shallow lattice, the Bloch dispersion curve is very close to that of free particles, except for momenta near the edge of the Brillouin zone $k=\pm k_G/2$, where the dispersion has a negative curvature (see, e.g., Ref.~\cite{ziman72:book}). Hence, the condition for the dynamical instability takes the form $k_G/2=mv/\hbar$. This coincides with the above condition for the break point, obtained within the generalized Landau criterion of superfluidity, suggested in Sec.~\ref{sec:general}. A similar analysis can also be carried out for the TG gas.

The frictionless motion at some values of the parameters $v$, $\gamma$, $\alpha$ is consistent with the presence of persistent currents in the 1D Bose-Hubbard model \cite{trombettoni01, altman05, polkovnikov05, ruostekoski05, kolovsky06, danshita09}. As discussed in Sec.~\ref{sec:general}, the drag force can be considered as a measure for superfluidity in the absence of the order paramenter. Then Fig.~\ref{fig:vgam} represents the phase diagrams in the $v$-$\gamma$ and $v$-$1/\alpha$ planes. They are similar to that of Ref.~\cite{polkovnikov05}. One can see from the diagrams that there is no sharp transition from superfluid to isolated phase in 1D, which is consistent with the experimental findings of Ref.~\cite{mun07}. Note that in paper \cite{polkovnikov05}, superfluidity was examined in terms of quantum phase slips, discussed in Sec.~\ref{sec:slip}. So, the both quasiparticle and quantum phase slip descriptions lead to the same results.

\subsection{1D Bose gas in a moving random potential}
\label{sec:random}

\begin{figure}[tbhp]
\noindent\includegraphics[width=.8\columnwidth,clip=true]{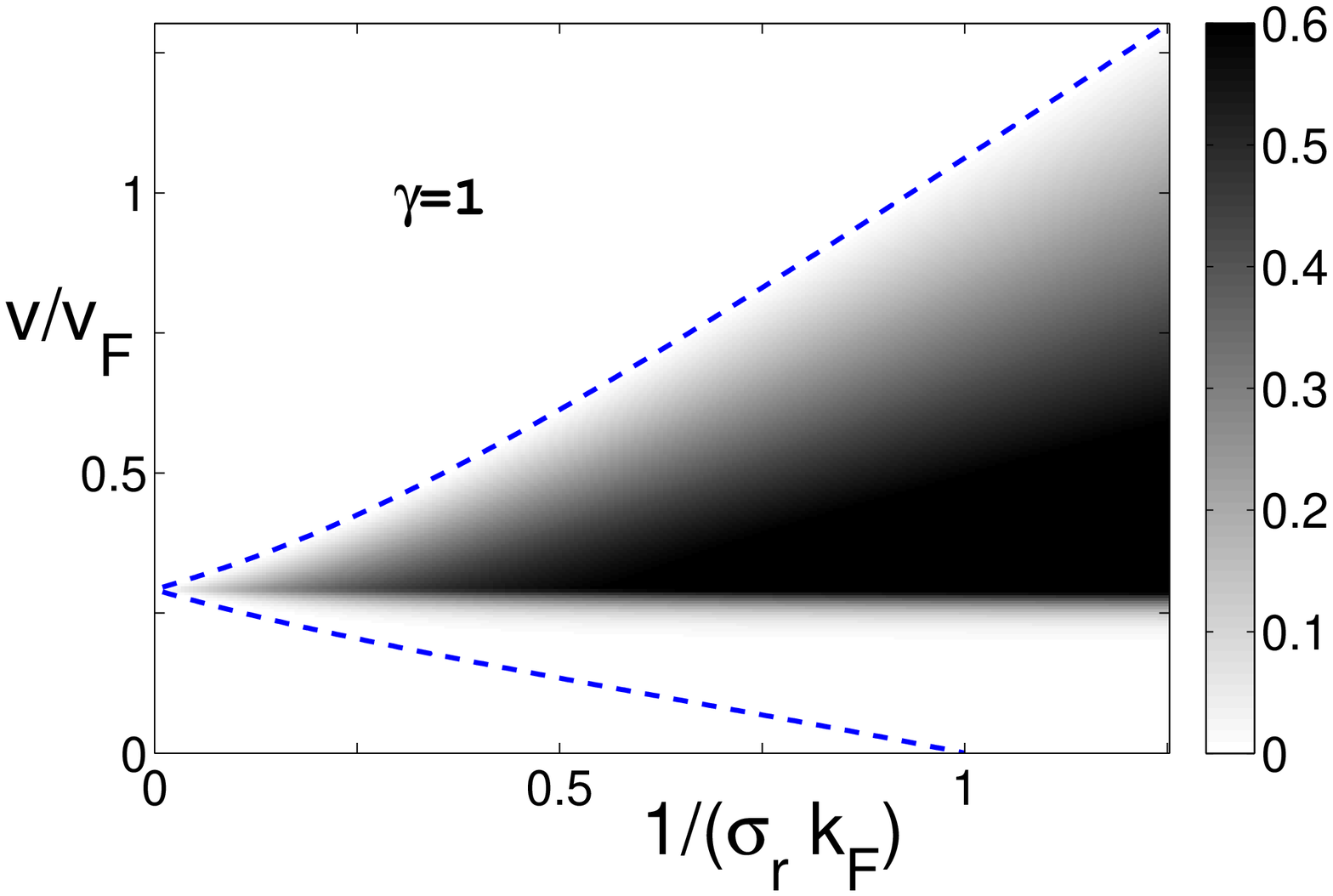}\\
\includegraphics[width=.8\columnwidth,clip=true]{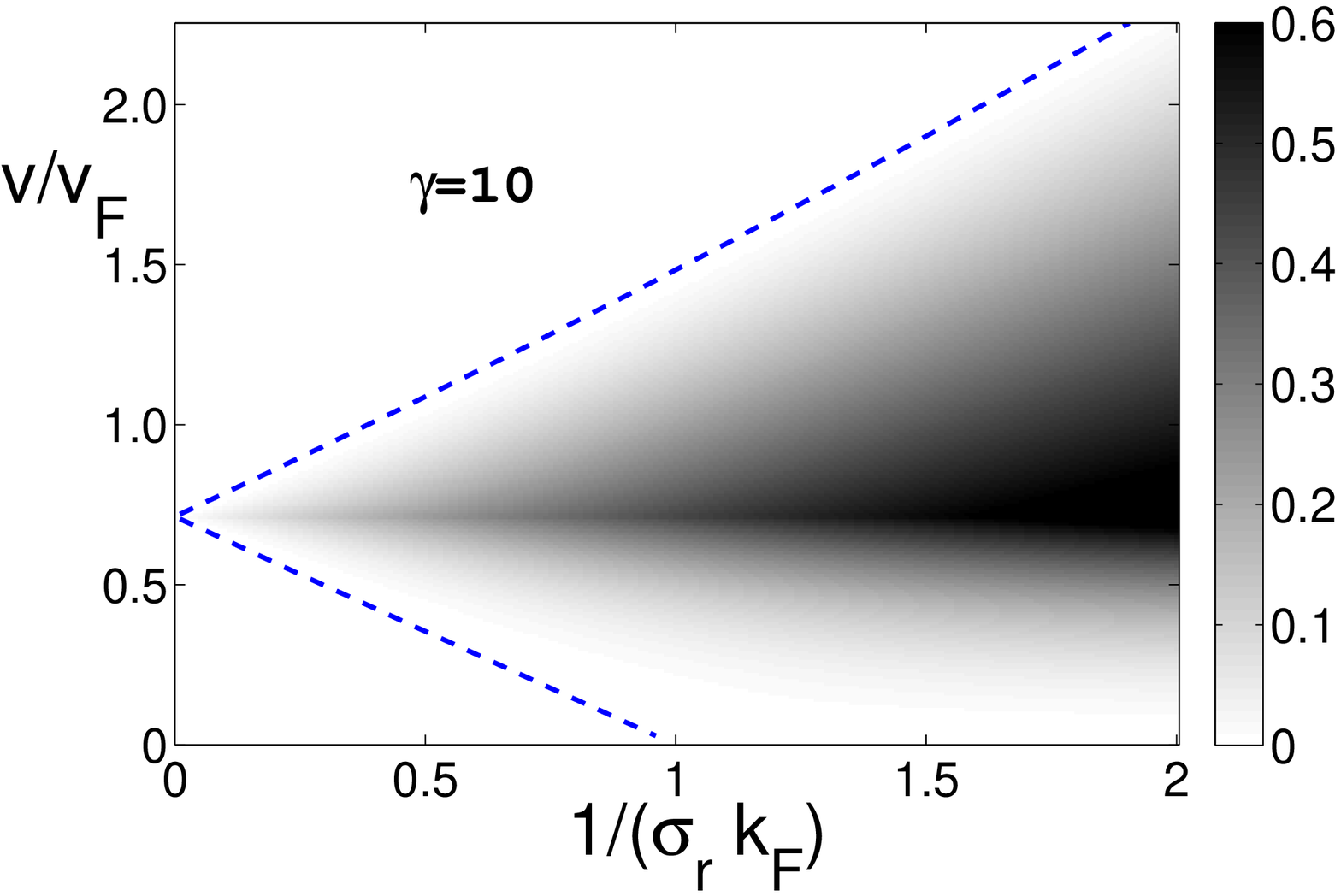}
\caption{\label{fig:vsig} (Color online) Zero temperature phase diagram for superfluid-isolator transition of the Bose gas in a moving random potential: drag force (in units of  $2\pi m V_\mathrm{R} ^2\sigma _r N/\hbar ^2$) versus the potential velocity (in units $v_F$) and the inverse correlation length (in units $\kf$). The dashed (blue) lines correspond to the intersection of the limiting dispersions $\omega_\pm$ and the transition line: $v=\sigma_\mathrm{r}\omega_{\pm}(2/\sigma_\mathrm{r})/2$. On the left to the lines, the drag force is zero, because within the integration limits of Eq.~(\ref{dfrand}), the values of the DSF is zero (see Fig.~\ref{fig:omplmi}).}
\end{figure}

Recently, direct observation of Anderson localization with a controlled disorder was reported in a one-dimensional waveguide \cite{billy08}. An initially trapped Bose gas is released into a one-dimensional waveguide, where it is exposed to an artificially created speckled potential. The wave function of the spreading atoms can be directly measured by optical methods.

Anderson localization was initially predicted for a system of non-interacting particles \cite{anderson58}. Modern experimental technique in cold atoms allows to control the strength of interparticle interactions and the parameters of speckled potentials as well (see, e.g., reviews \cite{sanchez10:rev,modugno10:rev}). Delocalization of a disordered bosonic system by repulsive interactions was observed \cite{deissler10}. Thus, the interplay between interactions and disorder remains a fundamental problem in this field. Considerable attention has recently been given to this problem in many papers \cite{sanchez06, sanchez07, lugan07, lugan07a, paul07, pikovsky08, kopidakis08, falco09, aleiner10,radic10}. Most of the papers consider weakly-interacting bosons or the regime of infinite interactions (TG gas).
By using the generalized Landau criterion of superfluidity, introduced in Sec.~\ref{sec:general}), one can study the behaviour of a 1D Bose gas in weak random potentials and obtain the superfluid-insulator phase diagram for \emph{arbitrary} strength of boson interactions.

It is convenient to consider random potentials by introducing an ensemble of various potentials. Then the random potential features can be obtained by averaging over the ensemble. One of the most important characteristics of random potentials, created with speckled laser beams, is their correlation function $\langle V_i(x)V_i(x')\rangle=g(x - x')$, where $\langle\cdots\rangle$ stands for the average over the random potential ensemble.
For an arbitrary potential profile, the drag force is calculated with Eq.~(\ref{dragf1D}). Taking the average of this equation with respect to the ensemble, we obtain the drag force acting from the moving random potentials. At zero temperature, we derive
\begin{equation}
\langle F_\mathrm{v}\rangle  = \int_0^{2k_C} \d k\,k \widetilde{g}(k)S(k,k v).
\label{dfrand}
\end{equation}
Here $\widetilde{g}(k)\equiv\langle |\widetilde{V}_{\mathrm{i}}(k)|^2\rangle/L$ is the Fourier transform of the correlation function $g(x)$. The integration in (\ref{dfrand}) is limited, because for speckled laser beams, the function $\widetilde{g}(k)$ always has a finite support due to the limited aperture of the diffusion plate generating the random phase \cite{goodman75:book,clement06}. So, $\widetilde{g}(k) = 0$ for $|k| > 2k_C$. For estimations, we take a realistic correlation function \cite{goodman75:book,clement06} $\widetilde{g}(k) = \pi V_\mathrm{R} ^2\sigma_\mathrm{r}\mathrm{\Theta}\!\! \left( 1 - \frac{|k|\sigma_\mathrm{r}}{2}\right)\left( 1 - \frac{|k|\sigma_\mathrm{r}}{2}\right)$. Here $\sigma _r\equiv 1/k_C$ is the random potential correlation length,
depending of the parameters of the experimental device. It is usually of order of 1 $\mu \mathrm{m}$~\cite{modugno10:rev}. Thus, we have three parameters governing the phase diagram: the potential velocity, the interaction strengh, and the correlation length.

Like the lattice potential, the random potential is non-local, and, as a consequence, the drag force (\ref{dfrand}) is proportional to the total number of particles. The results obtained with the interpolating formula (\ref{dsfapp1}) are shown in Fig.~\ref{fig:vsig}. As for the shallow lattice potentials, the figure can be treated as the zero-temperature phase diagram for the superfluid--insulator transion: superfluidity assumes zero or strongly suppressed values of the drag force.

In the Bogoliubov and TG regimes, one can find the drag force analytically. For $\gamma\ll 1$, we obtain from Eqs.~(\ref{dsfbog}) and (\ref{dfrand})
\begin{equation}
\langle F_{\mathrm{v}}\rangle  = F_0\,\Theta(\xi-\xi_c)
\Theta(1-t)(1-t).
\label{dfrandBog}
\end{equation}
Here  $F_0\equiv{2\pi m V_\mathrm{R} ^2\sigma _r N}/{\hbar ^2}$ is a unit of force, $t\equiv\pi n \sigma_\mathrm{r}\sqrt{\xi^2-\xi_c^2}$, and $\xi_c=\sqrt{\gamma}/\pi$ is the sound velocity (in units of $v_\mathrm{F}$) in the Bogoliubov regime. In the TG regime, Eqs.~(\ref{DSFTG}) and (\ref{dfrand}) yield
\begin{equation}
\langle F_{\mathrm{v}}\rangle  = F_0[f_1+(f_2-f_1)\Theta(\lambda_+ -\lambda_0)-f_2\Theta(\lambda_- -\lambda_0)],
\label{dfrandTG}
\end{equation}
where we introduce the notations $\lambda_0\equiv2/(\pi n \sigma_\mathrm{r})$, $\lambda_\pm\equiv2|\xi\pm 1|$, $f_1\equiv\frac{1}{4}(\lambda_+-\lambda_-)(1-\frac{\lambda_+ +\lambda_-}{2\lambda_0})$, $f_2\equiv\frac{(\lambda_0-\lambda_-)^2}{8\lambda_0}$.

The obtained results are in accordance with the existence of a mobility edge for a particle moving in a random potential with a finite correlation length $\sigma_\mathrm{r}$. In this case, the mobility edge is given by \cite{sanchez07,lugan07a} $k_\mathrm{mob}=k_C\equiv1/\sigma_\mathrm{r}$. If $|k|>k_C$, then the $k$-wave propagation is not suppressed,  while in the opposite case, the particle wave function is localized (Anderson localization), which leads to the particle immobility. In the TG regime, the gas is equivalent to the ideal Fermi gas, where the mobile particles lie in the vicinity of the Fermi points. In the reference frame where the random potential is at rest, the absolute value of momentum of the lowest Fermi level is given by $\hbar\kf'=|\hbar \kf - mv|$. When $\kf'>k_C$, the system should be superfluid. This is consistent with Eq.~(\ref{dfrand}), which yields zero value of the drag force for $\kf'>k_C$. Indeed, the value $2\kf'$ corresponds to the cross point of the dispersion curves for the TG gas and the transition line: $\omega_\pm(2\kf')=2\kf' v$, where $\omega_\pm$ is given by Eq.~(\ref{ompmstrong}) with infinite $\gamma$. Once $2\kf'> 2k_C$ then the value of drag force is zero, because the DSF is zero within the limits of integration in Eq.~(\ref{dfrand}), see Fig.~\ref{fig:omplmi}.

\section{Conclusion}
\label{sec:conclusion}

The absence of a well-defined order parameter makes the behavior of the 1D Bose gas rather unusual in comparison to the 3D case. As shown in Sec.~\ref{sec:landau}, the 1D Bose gas exhibits superfluid phenomena of equilibrium type (Hess-Fairbank effect, analogous to the Meissner effect in superconductivity) but in general does not show dynamic superfluid phenomena, such as persistent currents in a ring.
Instead of a phase transition to full superfluidity as it is known in 3D, the 1D Bose gas shows a smooth crossover and reaches the metastability of currents only in a weakly-interacting limit.
In this case, the drag force, being a simple integral parameter, can be chosen as \emph{a quantitative measure of superfluidity}. Superfluidity assumes zero or strongly suppressed value of the drag force for the Bose gas moving in different small external potentials. Which value of the drag force should be taken as a the transition threshold becomes a question of convention in 1D.

The drag force turns out to be fundamental, because it generalizes the Landau criterion of superfluidity. The generalized Landau criterion, based only on energy and momentum conservation, does work when the usual Landau criterion fails. This is because the drag force effectively involves not only the spectrum but \emph{the probability of transitions} to excited states. A good example is the dynamical instability of weakly-interacting 1D bosons, moving in a shallow lattice. As shown in Sec.~\ref{sec:shallow_lattice}, the generalized Landau criterion not only successfully predicts the dynamical instability of the system but gives the quantitative characteristics of the phase transition.

It should be noted that the suggested approach has an apparent disadvantage. Being based on first order time-dependent perturbation theory, the scheme cannot describe changes of the ground state caused by the perturbing potential. Despite of this fact, it can describe well the superfluid--insulator phase diagram, when the system propagates through shallow lattices or random potentials.

\begin{acknowledgments}
The authors are grateful to Antony Leggett, Lev Pitaevskii,  Hans-Peter B\"uchler, Andrey Kolovsky, Anatoly Polkovnikov, and Andrew Sykes for fruitful discussions.

J-SC acknowledges support from the FOM foundation of the Netherlands.  AYC acknowledges a visiting grant from the INSTANS activity of the European Science Foundation and the support from the JINR--IFIN-HH projects and thanks Massey University for hospitality. JB was supported by the Marsden Fund Council (contract MAU0706) from Government funding, administered by the Royal Society of New Zealand.
\end{acknowledgments}

\appendix
\section{General formula for the drag force from Fermi's golden rule}
\label{sec:dffgr}

Once an impurity moves into a homogeneous medium with velocity $\bm{v}$, it is scattered by the medium particles. In general, the scattering leads to transitions with momentum and energy transfer and, consequently, to a finite value of energy loss per unit time. It can be calculated within linear response theory \cite{pines66:book,pitaevskii03:book,astrakharchik04} or from Fermi's golden rule \cite{pines66:book,timmermans98}.

During the scattering, the initial state of the composite system is assumed to be $|\bm{k}_\mathrm{in},m\rangle =|\bm{k}_\mathrm{in}\rangle |m\rangle$, where $|m\rangle$ is initial state of the medium, the incident particle state is the plane wave $|\bm{k}\rangle=\exp(i\bm{k}\cdot\bm{y})/ \sqrt{V}$ with wave vector $\bm{k}_\mathrm{in}$, and $V=L^D$ is the $D$-dimensional volume. The wave vector is connected to the initial velocity by the relation $\bm{v}=\hbar \bm{k}_\mathrm{in}/m_\mathrm{i}$. The same notations are adopted for the final state $|\bm{k}_\mathrm{f},n\rangle= |\bm{k}_\mathrm{f}\rangle |n\rangle$. The rate for the scattering process (that is, the transition probability per unit time) is given in the lowest order in the impurity interaction by Fermi's golden rule
\begin{align}
w(\bm{k}_\mathrm{in},\bm{k}_\mathrm{f})=&\frac{2\pi}{\hbar}\sum_n
\big|\langle\bm{k}_\mathrm{f},n|H'|\bm{k}_\mathrm{in},m\rangle\big|^2\nonumber\\
&\times\delta(E_n+T_{\bm{k}_\mathrm{f}}-E_m-T_{\bm{k}_\mathrm{in}}),
\label{fgr}
\end{align}
where $T_{\bm{k}}=\hbar^2k^2/(2m)$ is the energy dispersion for the impurity. The momentum and energy transfer are given by
\begin{align}
\hbar\bm{q}&= \hbar\bm{k}_\mathrm{in} -\hbar\bm{k}_\mathrm{f}, \label{qtr}\\
\hbar\omega&=T_{\bm{k}_\mathrm{in}}-T_{\bm{k}_\mathrm{f}}=\hbar \bm{q}\cdot\bm{v}-\hbar^2 q^2/(2 m_\mathrm{i}),
\label{omtr}
\end{align}
respectively.

Performing the integration over $\bm{y}$ in the matrix element of the interaction Hamiltonian
$H'=\sum_j V_\mathrm{i}(|\bm{y}-\bm{x}_j|)$ yields for $\bm{q}\not=0$
\begin{align}
\big|\langle\bm{k}_\mathrm{f},n|H'|\bm{k}_\mathrm{in},m\rangle\big|^2 =|\widetilde{V}_\mathrm{i}(q)|^2\,
\big|\langle n|\delta\hat{\rho}_{\bm{q}}|m\rangle\big|^2/V^2.
\label{matrel}
\end{align}
At nonzero temperature, the medium can be in an arbitrary initial state $|m\rangle$ with the statistical probability $\exp(-\beta E_m)/{\cal Z}$. In this case, the transition rate (\ref{fgr}) should be averaged over the statistical ensemble, and we arrive at
\begin{equation}
w(\bm{q})=\frac{2\pi}{\hbar} \frac{|\widetilde{V}_\mathrm{i}(q)|^2}{V^2}
S\Big(q,\hbar \bm{q}\cdot\bm{v}-\frac{\hbar^2 q^2}{2 m_\mathrm{i}}\Big).
\label{wq}
\end{equation}
Here we use Eq.~(\ref{matrel}) and the definition of the DSF (\ref{sqomega}).

In order to obtain the energy loss per unit time, we need to sum up the energy transfer (\ref{omtr}) weighted with the rate (\ref{wq}) over all the final states
\begin{equation}
\dot{E}=-\sum_{\bm{q}} w(\bm{q})\Big(\hbar \bm{q}\cdot\bm{v}-\frac{\hbar^2 q^2}{2 m_\mathrm{i}}\Big).
\label{eloss1}
\end{equation}
Replacing the sum by the integral in the thermodynamic limit $\sum_{\bm{q}} \to V/(2\pi)^D\int \d^D q$ and subsituting Eq.~(\ref{wq}) yield Eq.~(\ref{enloss}).

\section{The drag force in the RPA approximation}
\label{app:DF_RPA}

\subsection{The general expression}

In this appendix, we will use the dimensionless variables $\lambda\equiv q/\kf$, $\nu\equiv\hbar\omega/\ef$, $\xi\equiv v/v_{\mathrm{F}}$, where $\kf\equiv\pi n$, $\ef\equiv\hbar^{2}\kf^{2}/(2m)$, and $v_{\mathrm{F}}\equiv\hbar\kf/m$ are the Fermi wave vector, energy, and velocity, respectively. Besides, it is convenient to introduce the small parameter $\alpha\equiv 2/\gamma$, which can take the values $0\leqslant\alpha\leqslant 1/4$ within the RPA scheme \cite{cherny06}. In terms of the new variables, Eq.~(\ref{dragf1Dpoint}) reads
\begin{equation}
f_\mathrm{v}\equiv\frac{F_\mathrm{v}\pi\ef}{g_{\mathrm{i}}^{2}\kf^{3}}=
\int_{0}^{+\infty}\d \lambda\,\lambda\, s(\lambda,2\xi\lambda), \label{dragf1Ddimless}
\end{equation}
where we put $s(\lambda,\nu)\equiv\ef {S(\kf\lambda,\ef\nu/\hbar)}/{N}$.

\begin{figure}[tbhp]
\includegraphics[width=.8\columnwidth]{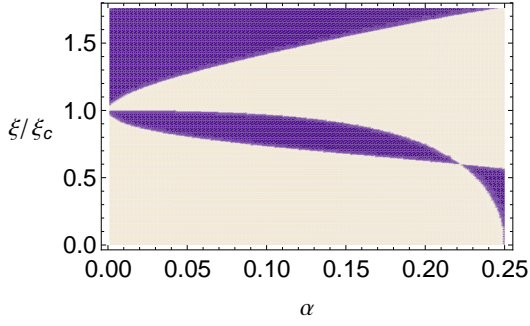} \caption{\label{fig:delta_contribution} (Color online) The dark (blue) region shows where the $\delta$-function contribution exists. Such a contribution results from the intersection of the curve $\nu_0(\lambda)$ and the line $2\xi\lambda$ in the $\nu$-$\lambda$ plane. One can see that there is no $\delta$-function contribution at sufficiently small velocity of the impurity. Here $\xi_{\mathrm{c}}\equiv c/v_{\mathrm{F}}$, and $\xi_{\mathrm{c}}\simeq 1-2\alpha$ in the strong-coupling regime. }
\end{figure}

The DSF in the RPA approximation was calculated and described in details in Sec. IVB of paper \cite{cherny06}. The result at zero temperature can be written in the form
\begin{equation}
s(\lambda,\nu)=s_{\mathrm{reg}}(\lambda,\nu)+\widetilde{A}(\lambda)\delta(\nu-\nu_0(\lambda))
\label{DSFRPA}
\end{equation}
with the regular part of the dimensionless DSF
\begin{equation}
s_{\mathrm{reg}}(\lambda,\nu)\equiv\frac{\lambda}{4}\frac{(1-3\alpha)^{2}(1-2\alpha)}
{\big[(1-3\alpha)^{2}\lambda-\alpha h\ln f \big]^{2}+[\alpha\pi h]^{2}}. \label{dsfreg}
\end{equation}
This part is localized in the same region $|\lambda^{2}-2\lambda|(1-2\alpha)\leqslant\nu\leqslant(\lambda^{2}+2\lambda)(1-2\alpha)$, as the DSF in the linear approximation, given by Eqs.~(\ref{DSFlinear}) and (\ref{ompmstrong}). Here we put by definition
\begin{equation}
f(\lambda,\nu)\equiv
\left|\frac{\nu^{2}-(\lambda^{2}-2\lambda)^{2}(1-2\alpha)^{2}}
{\nu^{2}-(\lambda^{2}+2\lambda)^{2}(1-2\alpha)^{2}}\right|,
\label{fdef}
\end{equation}
and
\begin{equation}
h(\lambda,\nu)\equiv(1-9\alpha/4)\frac{\lambda^{2}}{2} -\frac{\alpha}{2}
-\frac{\nu^{2}}{8\lambda^{2}}\frac{3\alpha(1-8\alpha/3)}{(1-2\alpha)^{2}}.
\label{heq}
\end{equation}
The quantities in Eq.~(\ref{DSFRPA}) are given by $\widetilde{A}(\lambda)\equiv A(\kf\lambda)/N$, $\nu_0(\lambda)\equiv\hbar\omega_0(\kf\lambda)/\ef$ with $A(q)$ and $\omega_0(q)$ being defined exactly in Sec. IVB of Ref.~\cite{cherny06}. By employing the general formula (\ref{dragf1Ddimless}) and the RPA expression (\ref{DSFRPA}), we can evaluate the drag force by numerical integration. The $\delta$-function contribution appears as a result of intersection of the curve $\nu_0(\lambda)$ and the line $2\xi\lambda$ in the $\nu$-$\lambda$ plane and does not play a role at small values of the impurity velocity, see Fig.~\ref{fig:delta_contribution}. One can easily show that the $\delta$-contribution appears for $\alpha\leqslant2/9$ when $\xi_{-}(\alpha)\leqslant\xi\leqslant \xi^{*}(\alpha)$ or $\xi\geqslant\xi_{+}(\alpha)$, where
\begin{align*}
&\xi^*(\alpha)= (1-2\alpha)\sqrt{
\frac{(1 - 3 \alpha)^2/\alpha^2 - 1}{3 (1 - 8\alpha/3) + (1 - 3\alpha)^2/\alpha^2}},\\
&\xi_{\pm}(\alpha)= (1-2\alpha)\frac{4 - 9\alpha \pm \sqrt{\alpha}\sqrt{16 - 71\alpha +
80\alpha^2}}{4(1 - 3\alpha + 2\alpha^2)}.
\end{align*}

\subsection{The drag force for small velocities at zero temperature}

Since at the small velocities the linear approximation (\ref{dragflinear}) fails, we need to apply here the full RPA expression for the drag force. At small values of $\xi=v/v_{\mathrm{F}}$, the limits of the integral in Eq.~(\ref{dragf1Ddimless}) becomes very close to each other, and we can put $f_\mathrm{v}= \int_{2(1-\xi/\xi_{\mathrm{c}})}^{2(1+\xi/\xi_{\mathrm{c}})}\d \lambda\,\lambda\, s(\lambda,\nu=2\xi\lambda) \simeq s_{\mathrm{reg}}(2,4\xi) 8\xi/\xi_{\mathrm{c}}$ and obtain from Eq.~(\ref{dsfreg})
\begin{equation}
f_\mathrm{v}\simeq\frac{4\xi(1-3\alpha)^{2}} { D^{2} +\alpha^{2}\pi^{2}
\Big[2-5\alpha-\frac{\xi^{2}}{\xi_{\mathrm{c}}^{2}}\frac{3\alpha}{2}(1-\frac{8\alpha}{3})\Big]^{2}
},
\label{dragfsmallv}
\end{equation}
where
\[
D\equiv
2(1-3\alpha)^{2}
+\alpha\bigg[2-5\alpha-\frac{\xi^{2}}{\xi_{\mathrm{c}}^{2}}\frac{3\alpha}{2}\Big(1-\frac{8\alpha}{3}\Big)\bigg]
\ln\Big(4\frac{\xi_{\mathrm{c}}^{2}}{\xi^{2}}-1\Big).
\]

In principle, the logarithmic term should dominate in the denominator when $\xi\to0$ and one can neglect all the other terms; however, this approximation works only at very small values of $\xi$. A much better approximation can be obtained by keeping the logarithmic term together with the zero-order terms in $\xi$:
\begin{equation}
f_\mathrm{v}\simeq\frac{(1-3\alpha)^{2}}{(1-5\alpha/2)^{2}}
\frac{\xi}{\Big[\frac{(1-3\alpha)^{2}}{(1-5\alpha/2)}
-2\alpha\ln\frac{\xi}{2\xi_{\mathrm{c}}}\Big]^{2}+\alpha^{2}\pi^{2}}.
\label{dragfsmallv1}
\end{equation}

\bibliography{df}

\end{document}